\documentclass[universe,article,submit,pdflatex,moreauthors]{Definitions/mdpi}
\pdfoutput=1
\firstpage{1} 
\pubvolume{1}
\issuenum{1}
\articlenumber{0}
\pubyear{2022}
\copyrightyear{2022}
\datereceived{} 
\dateaccepted{} 
\datepublished{} 
\hreflink{https://doi.org/} 

\usepackage{color}
\usepackage{mathtools}
\usepackage{mathrsfs}
\usepackage{calrsfs}
\usepackage{comment}

\usepackage{multirow} 
\preto{\abstractkeywords}{\nolinenumbers}

\newcommand{\lPl}{\ell_{_{\rm Pl}}}
\DeclareMathOperator{\Tr}{Tr}

\Title{Effective field theory description of horizon-fluid determines the scrambling time}
\TitleCitation{Effective field theory description of horizon-fluid determines the scrambling time}
\Author{Swastik Bhattacharya$^1$, and S. Shankaranarayanan$^2$*\orcidA{}}
\address{%
$^{1}$Department of Physics, BITS Pilani Hyderabad, Hyderabad 500078, Telangana State, India; swastik@hyderabad.bits-pilani.ac.in \\
$^{2}$ Department of Physics, Indian Institute of Technology Bombay, Mumbai~400076, India, shanki@iitb.ac.in}
\firstnote{These authors contributed equally to this work.}
\corres{Correspondence: shanki@iitb.ac.in}
\abstract{Black hole horizons interact with external fields when matter-energy falls through them. Such non-stationary black hole horizons can be described using viscous fluid equations. This work attempts to describe this process using effective field theory methods. Such a description can provide important insights beyond classical black hole physics. In this work, we construct a low-energy effective field theory description for the horizon fluid of a 4-dimensional, asymptotically flat, Einstein black hole. The effective field theory of the dynamical horizon has two ingredients: degrees of freedom involved in the interaction with external fields and symmetry. The dual requirements of incorporating near-horizon symmetries (${\cal S}1$ diffeomorphism) and possessing length scales due to external perturbations is naturally satisfied if the theory on the non-stationary black hole horizon is a deformed Conformal Field Theory (CFT). For the homogeneous external perturbations, at the lowest order, this leads to $(2 + 1)-$dimensional massive scalar field where the mass is related to the extent of the deformation of the  CFT. We determine the mass by obtaining the correlation function corresponding to the effective field and relating it to the bulk viscosity of the horizon fluid. Additionally, we show that the coefficient of bulk viscosity of the horizon fluid determines the time required for black holes to scramble. Furthermore, we argue that matter-field modes with energy less than $m_{\rm eff}$ falling into the horizon thermalize more slowly. Finally, we construct a microscopic toy model for the horizon fluid that reduces to the effective field theory with a single scalar degree of freedom. We then discuss the usefulness of the effective field model in understanding how information escapes from a black hole at late times.}

\begin{document}
\section{Introduction}

There are deep interconnections between gravity, quantum theory, and thermodynamics~\cite{2001-Wald-LRR,2003-Jacobson.Parentani-FP,2008-Carlip-Lec}. 
The laws of black hole mechanics describe the entropy and temperature of black holes, which are a consequence of quantum mechanics in the presence of strong gravity~\cite{1973-Bardeen.etal-CMP,1995-Jacobson-PRL,2010-Padmanabhan-RPP,2015-Padmanabhan-Entropy,1975-Hawking-CMP}.
Black hole thermodynamics points to two issues: First,  it demands a statistical mechanical origin of entropy. It has been argued that most of the black hole degrees of freedom (DOF) reside on the horizon, as the black hole entropy scales as area~\cite{2001-Wald-LRR,2007-Das.Shankaranarayanan-CQG}. Second, it deals only with equilibrium states.

Due to gravity, the black hole horizon continually interacts with external fields (perturbations) and is non-stationary. The interaction leads to an energy transfer from the external fields to black hole degrees of freedom. Formally, the action for such a description is:
\begin{equation}
S_{\rm Total} = S_{\rm BH} + S_{\rm Ext} + 
S_{\rm Int}[{\rm BH}, {\rm Ext}] 
\label{eq:Formalaction}
\end{equation}
where $S_{\rm Ext}(S_{\rm BH})$ is the action corresponding to external fields (isolated black hole), and $S_{\rm Int}$ corresponds to the interaction. The interaction term leads to \emph{dissipative} effects when computing observables involving black holes. In classical black hole physics, this is explicitly seen by projecting the equations of motion of the fields and gravity on the black hole event horizon leading to dissipative equations. In these scenarios, fluid dynamics description is helpful as only average quantities resulting from the interactions at the microscopic level are observed on macroscopic scales~\cite{2019-Romatschke.Romatschke-Book}. Interestingly, it was shown that the black hole horizon behaves like a viscous fluid and satisfies Damour-Navier-Stokes equation~\cite{1982-Damour-Proc,1986-Price.Thorne-PRD,1986-Thorne.etal-Membrane}. 

Ideally, one should describe this horizon-fluid starting from a fundamental theory of Quantum Gravity. Unfortunately, the conceptual and technical obstacles in formulating a consistent quantum theory of gravity are formidable. {In this context, one of the issues that arise is the full diffeo-invariance in the entire theory, also known as background independence. General relativity is distinguished by its diffeomorphism or reparametrization invariance. In our model, however, perturbative expansion around a background metric breaks diffeomorphism invariance.
However, a complete Quantum gravity theory requires nonperturbative techniques that explicitly ensure diffeomorphism invariance [For a more detailed discussion, see Refs.~\cite{2022-Tessarotto1,2022-Tessarotto2}. For all the different lattice approaches, the challenge consists in showing that a continuum limit exists for which the effective action for the metric is one of the Einstein-Hilbert types. [see, for instance, Ref.~\cite{2009-Hamber-GRG,2012-Wetterich-PRD}.]}

It is generally thought that the quantum gravity effects are relevant close to the singularity at the center of a black hole. Hence, black holes are used as theoretical laboratories to test quantum gravity models. As mentioned above, most of the effort in the literature has been to understand the microscopic origin of black hole entropy. However, black hole thermodynamics now has the {\sl problem of Universality}~\cite{2008-Carlip-Lec}; at the leading order, several approaches using completely different microscopic degrees of freedom lead to Bekenstein-Hawking entropy~\cite{2001-Wald-LRR}. It is currently impossible to identify the true degrees of freedom responsible for the black hole entropy~\cite{2003-Jacobson.Parentani-FP}. Therefore, other tests are key in distinguishing such models. 
While reproducing the black hole entropy is only one test of a microscopic model of quantum gravity, obtaining transport coefficients can help us choose between various scenarios. To our understanding, deriving transport coefficients from quantum gravity is only partially addressed. 

Using a phenomenological approach, we determined the coefficient of bulk-viscosity for asymptotically flat black holes in general relativity~\cite{2016-Bhattacharya.Shankaranarayanan-PRD,2017-Cropp.etal-PRD}. 
In this work, we deduce the coefficient of bulk-viscosity of the horizon-fluid from a low-energy effective field theory on the dynamical horizon. See sec. \ref{sec:EFTofHorizon} for the detailed effective field theory description.

As an offshoot of our approach, we obtain the thermalization rate of black holes as matter energy falls into them. The black hole is perturbed when external matter energy falls into a stationary black hole. When the black hole settles into a stationary state, the black hole entropy increases as given by the first law of black hole mechanics with a different temperature~\cite{2001-Wald-LRR}. This can be interpreted as the thermalization of the infalling matter at the event horizon for an outside observer. For the \emph{homogeneous perturbations} of the event horizon, we provide an understanding of the thermalization at the horizon by relating the scrambling time~\cite{2008-Sekino.Susskind-JHEP} with the bulk viscosity of the horizon fluid. Using the scaling relations for the effective scalar degree of freedom, we explicitly show that the rate of thermalization slows down
as the perturbed black hole approaches the stationary point. 

Finally, we propose a microscopic toy model for the horizon fluid --- a two-dimensional integrable lattice model. Specifically, we consider {\sl Eight-vertex Baxter model}~\cite{1971-Baxter-PRL,1973-Baxter.Wu-PRL,1976-Baxter-JStatPhys,1977-Baxter-JStatPhys,1978-Baxter-JStatPhys,2016-Baxter-Book} as a model to explain the effective scalar degree of freedom. The Eight-vertex Baxter model reduces to an effective field theory with a single scalar degree of freedom in the continuum limit. We demonstrate that this microscopic model incorporates all the features.

The rest of this article is organized as follows: In sec. \eqref{sec:EFTofHorizon} we obtain the effective field theory of the event horizon interacting with homogeneous external perturbations. Using the mean-field theory description of horizon-fluid, we relate the scalar field $\varphi$ with the physical quantity associated with the horizon. 
In sec. \eqref{sec:BulkViscosity}, we calculate the bulk viscosity coefficient ($\zeta$) of the horizon-fluid from the correlation functions of the effective field theory Hamiltonian \eqref{H_Th_Defn}. We use Jeon's procedure to calculate $\zeta$ of the horizon-fluid from the correlations of the field's energy-momentum tensor~\cite{1995-Jeon-PRD}.
In sec. \eqref{sec:Thermalization}, we show that the effective theory can be used to predict the thermalization rate of the infalling matter energy to the black hole. 
In sec. \eqref{sec:Toymodel}, we explicitly construct a microscopic model which satisfies the requirements of symmetry and dynamical degrees of freedom of the effective field theory. Moreover, we show that the microscopic model reduces to the effective field theory Hamiltonian in the continuum limit. 
Finally, in sec. \eqref{sec:conc}, we discuss the implications of our results. The three appendices contain the details of the calculations.
In this work, we use natural units; we set $\hbar = c = G = k_B = 1$.

\section{Effective field theory description of the dynamical horizon}
\label{sec:EFTofHorizon}

Like in any effective theory, the effective field theory of the dynamical horizon has two ingredients~\cite{2004-Burgess-LRR,2020-Penco-Arxiv}: Degrees of freedom and Symmetries. Since bulk viscosity appears as a change in the cross-section area of the black hole horizon~\cite{2016-Bhattacharya.Shankaranarayanan-PRD,2017-Cropp.etal-PRD}, the process can be described by an {effective scalar degree of freedom}. To understand this, let us consider the process of infalling matter energy into the black hole, which increases the black hole area. This process can be viewed as an increase in the entropy since the entropy of a black hole is proportional to its area modulo the correction terms. 
Indeed, focusing on the {homogeneous processes} by which matter-energy falls into a black hole and using a phenomenological description, we showed that the evolution equation for the event horizon of a black hole follows from a Langevin equation~\cite{2015-Bhattacharya.Shankaranarayanan-IJMPD,2018-Bhattacharya.Shankaranarayanan-IJMPD}. Thus, the effective field theory describing the homogeneous process involves only one effective scalar degree of freedom. 

To constrain the form of the effective action, we need to \emph{identify the symmetries}. Stationary, non-extremal black holes in 4-dimensional general relativity exhibit an infinite-dimensional symmetry in the near-horizon region~\cite{1998-Kaul.Majumdar-PLB,2011-Kaul.Majumdar-PRD,2001-Koga-PRD,2001-Hotta.etal-CQG,2002-Hotta-PRD,1999-Carlip-PRL,2011-Carlip-Entropy,2010-Barnich.Troessaert-PRL,2016-Donnay.etal-PRL}.  Thus, the near-horizon possess infinite-dimensional algebra such as $\mathcal{S}1$ diffeomorphism~\cite{1999-Carlip-PRL,2011-Carlip-Entropy} or (near) BMS~\cite{2016-Donnay.etal-PRL,1962-Bondi.etal-PRLSA,1962-Sachs-PRLSA,2017-Strominger-Arx}. 
It {has been argued} that the Conformal Field Theory (CFT) on the black hole horizon can partly incorporate the near-horizon symmetries.

The CFT describing a stationary black hole can incorporate the $\mathcal{S}1$ diffeosymmetry~\cite{1999-Carlip-PRL,2011-Carlip-Entropy} as it possesses a representation of the Virasoro algebra\footnote{The conformal anomaly gives rise to the central term in the Virasoro algebra.}~\cite{2011-Carlip-Entropy,2012-Compere-LRR,Averin_2019}. Hence, it is a natural candidate for a low-energy effective theory of stationary black holes~\cite{1998-Kaul.Majumdar-PLB,2011-Kaul.Majumdar-PRD,2001-Koga-PRD,2001-Hotta.etal-CQG,2002-Hotta-PRD,1999-Carlip-PRL,2011-Carlip-Entropy}. This is the viewpoint we shall adopt in this work. We demonstrate that even a simple, effective theory toy model constructed from this starting point can help us understand the transport properties of the horizon fluid. For our purposes here, this will be sufficient as the details of the phase space of the theory on the horizon do not concern us. 

A dynamical (non-stationary) black hole can be viewed as interacting with external fields. 
Physically, we can view this process as a perturbed black hole relaxing to a stationary black hole by emitting QNMs~\cite{1992-Chandrasekhar-BHBook}. Interacting horizon fluid with external fields leads to conformal symmetry breaking. Thus, within the effective field theory approach, this means adding interaction terms to stationary black holes described by CFT.

We proceed by incorporating the near-horizon symmetries for a perturbed CFT. The perturbed CFT we choose possesses symmetries that lead to a representation of the Virasoro algebra~\cite{1987-Zamolodchikov-JETPL}. These are \emph{Integrable field theories} with an infinite number of conserved charges corresponding to an infinite number of symmetries~\cite{1989-Zamolodchikov-Proc}. The crucial point that allows us to model the black hole horizon-fluid by such a perturbed CFT is that one of the representations of the Virasoro algebra corresponding to the perturbed CFT is also a representation of the $\mathcal{S}1$  diffeomorphism symmetry~\cite{1987-Zamolodchikov-JETPL}. Thus, the effective field theory corresponding to perturbed CFT has \emph{at least} one length scale, and only a bare minimum input from the black hole physics is required. 

\begin{figure}[!htb]
\vspace*{-20pt}
    \centering
    \includegraphics[width=0.70\linewidth]{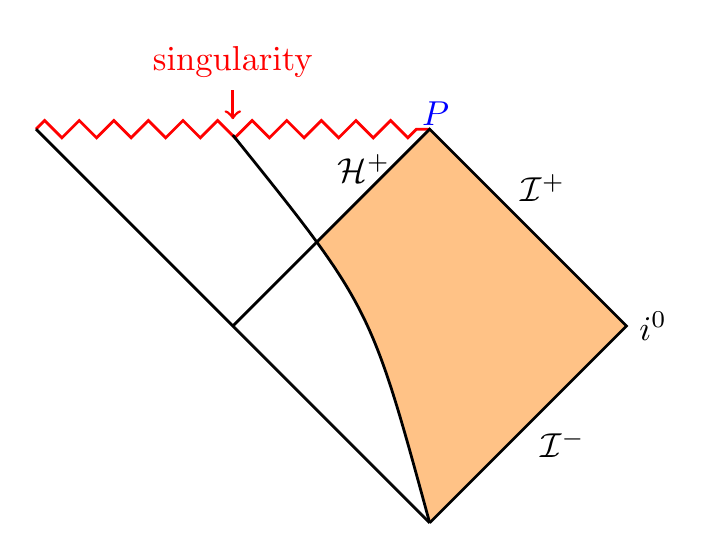}
\caption{Space-time diagram for the collapsing black hole.
The shaded region is the exterior of
a collapsing star, $r=0$ line at the top of the diagram is the
singularity. $\mathcal{H}^+$ denotes the event horizon. $P$ is the future timelike infinity.}
\label{fig:bhcollapse}
\end{figure}

The physical picture is the following: As shown in Fig. \ref{fig:bhcollapse}, we consider a collapsing black hole. The black hole is stationary only at point $P$ and is described by a CFT. A black hole that relaxes (from any point on $\mathcal{H}^+$) to the stationary black hole is described by deformed CFT. Thus, the effective field theory we use to model the dynamical black hole horizon here satisfies two requirements: First, at point $P$, it is described by a CFT. Second, the theory must reproduce the transport phenomena from any point on $\mathcal{H}^+$ to $P$. 
 
The simplest theory with a single scalar degree of freedom that is a deformed CFT and incorporates a representation of the $\mathcal{S}1$ diffeosymmetry is the \emph{free massive scalar field theory}. 
The corresponding Hamiltonian is 
\begin{equation}
H_{\rm eff }(\varphi) =   \int dA \,  \bigg[\frac{1}{2} \pi_{\varphi}^2 + \frac{1}{2}\big(\nabla\varphi\big)^2 +\frac{m_*^2}{2}\varphi^2\bigg] \, , \label{H_Th_Defn}
\end{equation}
where, $\pi_{\varphi} = {\partial \varphi}/{\partial t}$.
The above Hamiltonian is the minimal effective Hamiltonian regarding which we want to mention the following points: First, this is the Hamiltonian of the free scalar field ($\varphi$) in $(2+1)-$dimensional Minkowski space-time where $m_*$ is the mass of the excitations of the field $\varphi$ and the integral is over the area of the horizon. Interestingly, when the black hole becomes stationary, the effective scalar degree of freedom corresponds to a massless scalar field theory, and the theory becomes a CFT. Second, $m_*$ is related to the extent of the deformation of the CFT. Thus, $m_*$ contains all the information about the interaction between the isolated black hole and external fields in Eq.~\eqref{eq:Formalaction}. 

To obtain $m_*$, we need to identify the scalar degree of freedom  ($\varphi$) with a physical quantity associated with the horizon. Once we identify $\varphi$, we can use the above Hamiltonian to derive bulk viscosity and thermalization rate. In the rest of this section, we use Ginzburg-Landau formalism to identify the scalar field ($\varphi$) with a physical quantity associated with the horizon.
We achieve this by varying the Ginzburg-Landau entropy functional to determine the maximum value of entropy corresponding to the equilibrium state by carrying out the following steps:
\begin{enumerate}
    \item Associate the process of the perturbed black hole to a stationary black hole as a critical phenomenon.
    \item  Use Ginzburg-Landau formalism to phenomenological describe this process.
    \item Associate the Ginzburg-Landau functional to the entropy functional of the horizon-fluid~\cite{2016-Bhattacharya.Shankaranarayanan-PRD,2017-Cropp.etal-PRD}
    \item Identify the functional with 
    the effective scalar field Hamiltonian~\eqref{H_Th_Defn}.
\end{enumerate}

\subsection{Associate the process with critical phenomena}

As mentioned above, at the stationary point ($P$ in Fig.~\ref{fig:bhcollapse}), the black hole horizon is described by a CFT. In this framework, $P$ is the critical point, and the effective scalar field ($\varphi$) is massless $(m_* = 0)$. At any point on ${\cal H}^+$ in Fig.~\eqref{fig:bhcollapse}, $m_*$ is non-zero and relates the extent of the deformation of the CFT. Thus, as the non-stationary black hole settles down to the stationary state, $m_*$ flows from a non-zero value to zero at the critical point. Although other higher-order terms corresponding to deformed CFT can be present, $m_*^2 \varphi^2$ should always be present in any effective scalar field theory representing deformed CFT. Thus, the effective scalar field ($\varphi$) is associated with a phenomenological order parameter.

\subsection{Phenomenological description of the process using Ginzburg-Landau formalism}

Within the Ginzburg-Landau formalism, for many physical systems close to a critical point, it is possible to establish a phenomenological explanation of the increase in area/entropy caused by homogeneous perturbations~\cite{1990-Penrose.Fife-PhysicaD}. Identifying the Ginzburg-Landau functional with the scalar field Hamiltonian \eqref{H_Th_Defn} will allow us to connect the effective scalar field $\varphi$ explicitly with a black hole parameter. 

For the homogeneous perturbations of the black hole horizon, the Ginzburg-Landau functional can be written in terms of a \emph{single scalar order-parameter} ($\eta$). A natural choice for the Landau-Ginzburg functional is the $Z2$ symmetry-breaking terms. Since the homogeneous perturbations change the area of the black hole, it is natural to relate this scalar order parameter to the black hole horizon area. Like in phase-field models, we use entropy functional instead of energy functional, whose negative always decreases on solution paths~\cite{1990-Penrose.Fife-PhysicaD}. 
{There are two main factors: First, we investigate a scenario in which the black hole (of horizon area $A$) interacts with its surroundings, resulting in an energy flow. Therefore, we require a framework in which the energy density and the order parameter are treated on the same footing. As demonstrated in Ref. \cite{1990-Penrose.Fife-PhysicaD}, the relevant thermodynamic potential is entropy functional and not free energy functional. The stationary state of the black hole, indicated by the point $P$ in \ref{fig:bhcollapse}, corresponds to the thermodynamic equilibrium state of the black hole with maximum entropy. The deformed CFT corresponds to the quasi-stationary black hole (a point on $\mathcal{H}+$ other than $P$). Second, as we demonstrate, the analysis eliminates the arbitrariness in introducing the infra-red cut-off~\cite{2016-Bhattacharya.Shankaranarayanan-PRD}.}

\subsection{Associate the Ginzburg-Landau functional to the entropy functional of the horizon-fluid}
Using the phenomenological approach, the current authors have shown that modeling horizon-fluid as a critical system can provide a way to understand the black hole micro-states from the microscopic degrees of freedom of the horizon-fluid~\cite{2016-Bhattacharya.Shankaranarayanan-PRD,2016-Lopez.etal-PRD,2017-Bhattacharya.Shankaranarayanan-CQG,2017-Cropp.etal-PRD}. 
More specifically,  the local minimum value (or maximum value in the case of the entropy functional) corresponds to the equilibrium value of the entropy of the stationary black hole. 
Specifically, using Ginzburg-Landau formalism, it was shown that the order parameter of the homogeneous horizon-fluid is~\cite{2016-Bhattacharya.Shankaranarayanan-PRD,2017-Cropp.etal-PRD}:
\begin{equation}
\eta= C\sqrt{\cal A} \, ,
\label{eq:eta-def}
\end{equation}
where $C$ is a \emph{dimensionless constant} whose value can be fixed by relating to a macroscopic quantity and 
${\cal A}$ is a macroscopically small but finite element of the black hole horizon area that 
satisfies the condition ${\cal A}/A \ll 1$.
While the phenomenological approach allows us to construct an entropy functional, it is challenging to interpret the order parameter physically. However, once we identify the phenomenological energy functional with the effective Hamiltonian, the effective scalar field $\varphi$ can be interpreted as an order parameter.
The entropy functional (${\cal S}$) of this elemental area of the horizon-fluid about 
($T, {\cal A}$) is~\cite{1990-Penrose.Fife-PhysicaD}:
\begin{equation}
{\cal S} = {\cal S}_0(T, {\cal A}) - a \, \eta^2 - b\, \eta^4, \label{SMFT}
\end{equation}
where $a, b$ are constants. 
{${\cal S}_{QS}$ (${\cal S}_{S}$) represents the value of the entropy functional ${\cal S}$ in the quasi-stationary state (stationary state). ${\cal S}_S$ represents the global maximum for the entropy functional ${\cal S}$. Assuming the transition from ${\cal S}_{QS}$ to ${\cal S}_S$ is a slow physical process, equilibrium thermodynamics can be used to characterize the quasi-stationary state. For the effective field theory \eqref{H_Th_Defn}, this corresponds to the ground state of the deformed CFT steadily transitioning to the CFT state. Due to slow evolution, the deformed CFT vacuum is expected to possess some of the symmetries of a CFT state.} 

Rewriting (\ref{SMFT}) in terms of the horizon-area at equilibrium and using Eq. \eqref{eq:eta-def}, we get,
\begin{eqnarray}
S_{\rm max}= \frac{{\cal A}_{\rm max}}{4}= {\cal S}_0 -  a \,  C^2 {\cal A}_{\rm max} - b \, C^4  {\cal A}_{\rm max}^2 
&\mbox{where}& 
a= -\frac{1}{4C^2}; ~ {\cal S}_0- b\, C^4 \, {\cal A}^2= 0 \,.~
\label{a}
\end{eqnarray}
The change in the entropy functional is related to the difference in the energy density of the horizon-fluid:
\begin{equation}
\delta {\cal H}(\delta\eta) = -T \, \delta {\cal S}  = \frac{T}{2C^2}\delta\eta^2.
\label{H_eff_eta}
\end{equation}
where we have set $k_B = 1$. Expansion around the maxima implies that the terms proportional to $\delta\eta$ cancel leading to the condition $2b\eta^2_{max} = -a$. Substituting this in the terms quadratic in $\delta\eta$, (i. e. $(a+2b\eta_{max}^2)\delta\eta^2)$ leads to the above expression.
[The deviation of the variable from equilibrium is prefixed $\delta$.]

\subsection{Identify the functional with the effective scalar field Hamiltonian}

Associating the change in the energy density of the horizon-fluid [$\delta {\cal H}(\delta\eta)$] with 
$H_{\rm eff}(\varphi)$ \eqref{H_Th_Defn} leads to:
\begin{equation}
\int dA \, \delta {\cal H} \equiv \delta H =  \delta H_{\rm eff} \, .
\label{eq:HF-Heff}
\end{equation}
The above relation provides one easy route to relate the order parameter ($\eta$) with the effective scalar field ($\varphi$)
thus providing a way to understand horizon dynamics. Using the above relation, we can express the scalar field $\varphi$ in terms of the order parameter:
\begin{equation}
  \varphi=\sqrt{T} \frac{\delta\eta}{\sqrt{\cal A}} \equiv \sqrt{T} \tilde{\varphi}
  \label{eq:phi-tildephi-relation}
\end{equation}
Physically, ${\delta\eta}/{\sqrt{\cal A}}$ corresponds to the excess entropy density. Hence, $\tilde{\varphi}$ can be viewed as an order parameter for the entropy functional (${\cal S}$). 
For the homogeneous process, we can ignore the spatial gradient terms. Likewise, the kinetic term can be ignored since the field varies slowly. 
Thus, the change in the effective Hamiltonian \eqref{H_Th_Defn} for this process is:
\begin{equation}
\delta H_{\rm eff} =  \frac{T}{2} \, \int dA \, m_{*}^2 \, \tilde{\varphi}^2 \, , \quad 
\mbox{where}  \quad m_{*}= \frac{1}{\lPl\, C} 
\label{m_eff}
\end{equation}
where $\lPl$ is the Planck length which we set to unity and $C$ is dimensionless constant defined in \eqref{eq:eta-def}. 
Thus, we have related the effective scalar field Hamiltonian \eqref{H_Th_Defn} with black holes via the horizon-fluid. All the essential horizon physics is encoded in the Hamiltonian \eqref{H_Th_Defn} and the scalar field \eqref{eq:phi-tildephi-relation}. With this, we can now evaluate the correlation for this process from the correlation of the energy-momentum tensor of the effective field~\cite{1995-Jeon-PRD,Peskin:1995ev}. 

\section{Bulk viscosity from effective field theory} 
\label{sec:BulkViscosity}

In the previous section, 
we argued that $(2 + 1)-$dimensional massive free scalar field is the minimal effective field theory to describe the dynamical horizon with the mass ($m_{*}$) determining the extent of the deformation of the CFT. Using the relation between the effective field theory Hamiltonian \eqref{H_Th_Defn} and the horizon-fluid energy density, we associated $\varphi$ with the order parameter $\delta \eta$.
In this section, we calculate the bulk viscosity coefficient ($\zeta$) of the horizon-fluid from the correlation functions of the Hamiltonian \eqref{H_Th_Defn}. 
The fluctuation-dissipation theorem applies to these perturbations in the linear-regime~\cite{1966-Kubo-RPP}.
We use Jeon's procedure to calculate $\zeta$ of the horizon-fluid from the correlations of the field's energy-momentum tensor~\cite{1995-Jeon-PRD}. 
However, we need to make suitable changes in the formulation in Ref.~\cite{1995-Jeon-PRD} to apply to horizon fluid. To do this, we carry out the following steps: 
\begin{enumerate}
\item Obtain energy-momentum stress-tensor for the effective scalar field $\varphi$.
\item Modify Jeon's procedure~\cite{1995-Jeon-PRD} for the horizon-fluid. 
\item Obtain the correlation function corresponding to the homogeneous perturbations. 
\item Fix the constant by mapping to the macroscopic physics~\cite{1982-Damour-Proc}.
\end{enumerate}

\subsection{Obtain energy-momentum stress-tensor for the effective scalar field action}

The Lagrangian density corresponding to the effective Hamiltonian \eqref{H_Th_Defn} is,
\begin{equation}
\mathcal{L}_{\rm eff}= \frac{1}{2} \left[\dot{\varphi}^2 -(\nabla\varphi)^2\right]
- \frac{1}{2}m_*^2\varphi^2 
= \frac{1}{2}\big(\partial_{\mu}\varphi\partial^{\mu}\varphi 
- m_*^2\varphi^2) \, .
\label{Ldensity}
\end{equation}
The energy-momentum tensor ($T_{\mu\nu}$) of the effective scalar field ($\varphi)$ is:
\begin{equation}
T_{\mu\nu} = \frac{\partial \mathcal{L}_{\rm eff}}{\partial(\partial^{\mu}\varphi)}\partial_{\nu}\varphi - \eta_{\mu\nu}\mathcal{L}_{\rm eff}. \label{Energy-momentum}
\end{equation}
The non-zero components of the energy-momentum tensor are:
\begin{equation}
T_{ii} = \frac{1}{2}[\dot{\varphi}^2+ 2 (\partial_i\varphi)^2 - 
\delta^{kl} \partial_k \varphi \partial_l \varphi -m_*^2\varphi^2],  
T_{00}= \frac{1}{2}[\dot{\varphi}^2+ (\nabla\varphi)^2+ m_*^2\varphi^2]. \label{T_00}
\end{equation}
Here, $i$ take value 1 and 2. Thus, the trace of the spatial part is 
\[
T^i_i = \dot{\varphi}^2- m_*^2 \, \varphi^2 \, . 
\]

\subsection{Modify Jeon's procedure for the horizon-fluid}

Unlike the normal fluid, the stress-tensor of the horizon-fluid vanishes as the infalling matter-energy reaches the horizon, and the horizon becomes quasi-stationary~\cite{1982-Damour-Proc,1986-Price.Thorne-PRD,1986-Thorne.etal-Membrane}. More specifically, when the matter reaches an equilibrium at a given temperature, the stress tensor of the horizon fluid is zero. Thus, the field-theoretic description of the horizon-fluid corresponds to the deviation of the energy-momentum tensor of the field ($T_{\mu\nu}^H$) 
from its average at the thermal state, which we denote as $\langle T_{\mu\nu}^H\rangle_{\rm S}$. In other words, 
\[
\delta T_{\mu\nu}^H= T_{\mu\nu}^H- \langle T_{\mu\nu}^H\rangle_{\rm S} \, .
\]
Physically, this corresponds to the state when the expectation value of the stress-energy tensor of the perturbed CFT on the horizon is the thermal average. We can determine $\delta T_{\mu\nu}$ by tracking the deviation of the field $\varphi$ from its average value at equilibrium state, i.e., 
\begin{equation}
\varphi= \langle\varphi\rangle_{\rm S} + \delta\varphi
\end{equation}
where $\langle\rangle_{\rm S} \equiv \varphi_{\rm S}$ denotes the ensemble average of the density matrix.
{The equilibrium value of a physical quantity is determined by the thermal density matrix at temperature $T$.
Using the standard field theory techniques, we obtain: 
:}
\begin{equation}
\varphi_{\rm S}^2 = 
\frac{\sum e^{-\beta H} \varphi^2}{\sum e^{-\beta H}} 
= \frac{\int \mathcal{D}\varphi \, \varphi^2 e^{- \frac{\beta}{2} \int m_*^2 \varphi^2 d^2x}}{\int \mathcal{D}\varphi e^{- \frac{\beta}{2} \int m_*^2\varphi^2 d^2x}} = \frac{1}{2m_*^2}  \, .
 \label{phi_0}
\end{equation}

\subsection{Obtain the correlation function corresponding to the homogeneous perturbations}

Following Ref.~\cite{1995-Jeon-PRD}, the coefficient of bulk-viscosity ($\zeta$) of the horizon-fluid is:
\begin{equation}
\zeta = \frac{\beta}{2}\lim_{\omega\rightarrow 0}\lim_{{\mathbf q}\rightarrow 0}\sigma_{\bar{P}\bar{P}}, \label{zetasigma}
\end{equation}
where
\begin{eqnarray}
\label{PbarDef} 
\bar{P}(t,{\mathbf x}) &=&  P(t,{\mathbf x}) - v_S^2\rho(t,{\mathbf x})= \frac{1}{2}T_i^i(x^\mu)- v_S^2T_{00}(x^\mu),  \\
\label{sigma_P}
\sigma_{\bar{P}\bar{P}}(\omega,{\mathbf q}) &=& \frac{1}{2\pi {\cal A}} \int d^2{\mathbf x}\int_{-\infty}^{\infty}dt e^{-i{\mathbf q}.{\mathbf x}+i\omega t}\langle\bar{P}(t,{\mathbf x})\bar{P}(0)\rangle,  
 \end{eqnarray}
$2 \pi {\cal A}$ is the normalization constant for the spatial part, ${\cal A}$ is the area normalization and $v_S$ is the sound speed of the field which is $c$. Hence, it is set to unity. 
As mentioned above, in the case of horizon fluid, the relevant quantities are deviations of the energy-momentum tensor. Hence, the relevant quantity corresponding to the horizon-fluid in 
Eq. \eqref{PbarDef} is $\delta\bar{P}$.  
Thus, for the horizon fluid, 
Eqs.~\eqref{sigma_P} and \eqref{PbarDef} reduce to:
\begin{eqnarray}
\label{delta Pbar}
\delta\bar{P(t, {\mathbf x})} &=&  \frac{1}{2}\delta T_i^i(x^\mu)- \delta T_{00}(x^\mu) \\ 
\sigma_{\delta\bar{P},\delta\bar{P}}(\omega,{\mathbf q}) &=& 
 \frac{1}{2\pi {\cal A}} \int d^2{\mathbf x}\int_{-\infty}^{\infty}dt 
 e^{-i{\mathbf q} \cdot {\mathbf x}+i\omega t}\langle\delta\bar{P}(t,{\mathbf x})\delta\bar{P}(0)\rangle_{\rm S} \, . \label{sigmadelta}
\end{eqnarray}
From Eq.~\eqref{PbarDef}, we have, 
\begin{equation}
\bar{P}= -\frac{1}{2}(\nabla\varphi)^2- m_*^2\varphi^2. \label{PBar}
\end{equation}
In the case of homogeneous perturbations responsible for the bulk viscosity, we can ignore contributions from the $(\nabla\varphi)$ term. Physically, this corresponds to ignoring the contribution from the pole at ${\mathbf q} =0$. Thus, $\bar{P}= -m_*^2\varphi^2$. Rewriting $\varphi=\varphi_{\rm S}+\delta\varphi$, Eq.~\eqref{PBar} becomes:
\begin{equation}
\delta\bar{P}= -2m_*^2\varphi_{\rm S}\delta\varphi= -2\frac{m_*}{\sqrt{2}}\delta\varphi= \sqrt{2}m_*\delta\varphi= \frac{\sqrt{2}}{C}\delta\varphi \, , \label{DeltaP2}
\end{equation}
where we have used the relation Eq.~\eqref{phi_0}.
From the first-order time-dependent perturbation theory, the spectral density ($\rho_{\delta \bar{P} \, \delta \bar{P}}$) is given by~\cite{1995-Jeon-PRD}:
\begin{equation}
\rho_{\delta\bar{P} \delta\bar{P}}= \int d^3x e^{-i{\mathbf k}.{x}+i\omega t}\langle\big[\hat{A}_{\delta\bar{P}}(t,\mathbf{x}), \hat{A}_{\delta\bar{P}}(0)\big]\rangle_{\rm S} 
\, , \label{Rho}
\end{equation}
where $\hat{A}_{\delta\bar{P}}$ is the linear-response operator in the interaction picture. Due to the interaction, the ensemble average here is defined via the interaction Hamiltonian:
\begin{equation}
\hat{H_I}= \int d^2x F_{\delta\bar{P}}(t,{\mathbf x})\hat{A}_{\delta\bar{P}}(t,\mathbf{x}) \, , \label{H_I}
\end{equation}
where $F_{\delta\bar{P}}(t,{\mathbf x})$ is the generalised external force. 
Due to the teleological nature of the horizon~\cite{1986-Price.Thorne-PRD,2016-Bhattacharya.Shankaranarayanan-PRD}, we consider a process in which the external field is held constant for an extended period in the future (such that the system re-equilibrates in the presence of the external field):
\[
F_{\delta\bar{P}}(t,{\mathbf x})= F_{\delta\bar{P}}({\mathbf x})e^{\epsilon t}\theta(-t) \, ,
\]
where $\epsilon $ is an infinitesimal quantity and $\theta(-t)$ enforces the anti-casual nature of the horizon~\cite{1986-Price.Thorne-PRD,2016-Bhattacharya.Shankaranarayanan-PRD}. For details about the teleological condition, see Appendix \eqref{app:Teleo}. 
For the above external field, the spectral density is given by~\cite{1995-Jeon-PRD}: 
\begin{equation}
\rho_{\delta\bar{P} \delta\bar{P}}(\omega, {\mathbf q})=(1-e^{-\beta \omega})\sigma_{\delta\bar{P}\delta\bar{P}}(\omega, {\mathbf q}). \label{RhoSigma}
\end{equation}
Substituting the above expression in Eq.~\eqref{zetasigma}, we have:
\begin{equation}
\zeta = \frac{\beta}{2}\lim_{\omega\rightarrow 0}\lim_{{\mathbf q}\rightarrow 0}\frac{1}{1-e^{-\beta\omega}}\rho_{\delta\bar{P}\delta\bar{P}}(\omega, {\mathbf q}) \, , 
\label{ZetaLimit}
\end{equation}
which on substituting in Eq. \eqref{Rho} leads to:
\begin{equation}
\zeta = \frac{1}{2}\lim_{{\mathbf q}\rightarrow 0}\lim_{\omega\rightarrow 0}\frac{1}{\omega}\int d^3x e^{-i{\mathbf q}.{\mathbf x}+i\omega t}\langle\big[\hat{A}_{\delta\bar{P}}(t,\mathbf{x}), \hat{A}_{\delta\bar{P}}(0)\big]\rangle_{\rm S} \, . \label{Zeta3}
\end{equation}
Note that the relation between $\hat{A}_{\delta\bar{P}}$ and $\delta\bar{P}$ is still unknown. We can establish the relation using Eq.  \eqref{sigmadelta}. 
Specifically, multiplying Eq.  \eqref{sigmadelta}  with $(1-e^{-\beta\omega})$ and using the anti-casual, teleological nature of the horizon, we have:
\begin{equation}
(1-e^{-\beta\omega})\sigma_{\delta\bar{P}\delta\bar{P}}= \operatorname{Im}\bigg[\frac{1}{i}(1-e^{-\beta\omega})\int d^2xdt e^{-i{\mathbf q}.{\mathbf x}+i\omega t}\langle \delta\bar{P}(t,\mathbf{x})\delta\bar{P}(0)\rangle\bigg].
\end{equation}
Following Kubo~\cite{1966-Kubo-RPP}, the fluctuation-dissipation theorem allows us to express the RHS of the above equation as:
\begin{multline}
\operatorname{Im} \bigg[\frac{1}{i}(1-e^{-\beta\omega})\int d^2xdt e^{-i{\mathbf q}.{\mathbf x}+i\omega t}\langle \delta\bar{P}(t,\mathbf{x})\delta\bar{P}(0)\rangle\bigg] = \\
\operatorname{Im} \bigg[\int d^2x\int_{-\infty}^{\infty}dt \langle\big[\delta\bar{P}(t,\mathbf{x}), \delta\bar{P}(0)\big]\rangle e^{-i{\mathbf q}.{\mathbf x}+i\omega t}\bigg]. \label{FD}  
\end{multline}
Substituting the above expression in Eq.~\eqref{RhoSigma}, we have:
\begin{equation}
\rho_{\delta\bar{P}\delta\bar{P}}(\omega, {\mathbf q})= \int \int d^2x dt e^{-i{\mathbf q}.{\mathbf x}+i\omega t}\langle\big[\delta\bar{P}(t,\mathbf{x}), \delta\bar{P}(0)\big]\rangle . \label{Rhof}
\end{equation}
Comparing Eqs. \eqref{Rho} and \eqref{Rhof}, we identify $\hat{A}_{\delta\bar{P}}$ with $\delta\bar{P}(t,\mathbf{x})$. Thus, Eq.~\eqref{Zeta3} becomes:
\begin{equation}
\zeta = \frac{1}{4 \pi {\cal A}} \lim_{{\mathbf q}\rightarrow 0}\lim_{\omega\rightarrow 0}\frac{1}{\omega}\int d^3x e^{-i{\mathbf q}.{\mathbf x}+i\omega t}\langle\big[\delta\bar{P}(t,\mathbf{x}), \delta\bar{P}(0)\big]\rangle. \label{zeta4}
\end{equation}
Using the integral representation of the Theta function, 
\begin{equation}
\theta(-t)= - \lim_{\epsilon\rightarrow 0}\frac{1}{2\pi i}\int_{-\infty}^{\infty}\frac{e^{- ikx}}{x-i\epsilon}dx \qquad \epsilon > 0 \, , 
\label{ThetaFn}
\end{equation}
Eq.~\eqref{zeta4} can be expressed as an advanced Green's function, i. e.,
\begin{equation}
\zeta = 2\pi \operatorname{Im}  \Bigg[\frac{1}{4 \pi {\cal A}}\lim_{{\mathbf q}\rightarrow 0}\lim_{\omega\rightarrow 0}\int d^3x e^{-i{\mathbf q}.{\mathbf x}+i\omega t}\langle\big[\delta\bar{P}(t,\mathbf{x}), \delta\bar{P}(0)\big]\rangle\theta(-t)\Bigg]. 
\label{ZetaR}
\end{equation}
Substituting Eq.~\eqref{DeltaP2} in Eq.~\eqref{ZetaR}, one gets, 
\begin{equation}
\zeta = 2\pi\operatorname{Im} \Bigg[\lim_{{\mathbf q}\rightarrow 0}\lim_{\omega\rightarrow m_*}\frac{1}{C^2}\int d^3x e^{-i{\mathbf q}.{\mathbf x}+i\omega t}\langle\big[\delta\hat{\varphi}(t,\mathbf{x}),\delta\hat{\varphi}(0) \rangle\ \theta(-t) \Bigg]
. \label{Zeta5}
\end{equation}
$\langle\big[\delta\hat{\varphi}(t,\mathbf{x}),\delta\hat{\varphi}(0)\big]\rangle\theta(-t)$ is the advanced Green's function for the perturbed scalar field around its equilibrium. For homogeneous perturbations, the above expression reduces to: 
\begin{equation}
\zeta = \operatorname{Im} \Big[\lim_{\omega\rightarrow m_*}\frac{1}{C^2}\int_{-\infty}^{\infty} dt e^{i\omega t}\langle\big[\delta\hat{\varphi}(t),\delta\hat{\varphi}(0)\big]\rangle\theta(-t)\Big]. \label{ZetaF}
\end{equation}
Using the canonical quantization formalism discussed in Appendix \ref{App:QFTBulkViscosity}, we get:
\begin{equation}
\zeta[\omega_{\rm IR}] = 
\operatorname{Im}\left[ 
 -\frac{i}{4 C^2} \frac{\omega_{\rm IR}}{E_\beta(\omega_{\rm IR})} 
 \langle\hat{\varphi}^2_{\omega_{\rm IR}}(0)\rangle \right] \, .
\label{Zeta5-Final}
\end{equation}
where $E_\beta(\omega_{\rm IR})$ is  the average energy of excitation in the mode with frequency $\omega_{\rm IR}$ at
temperature $T$ and is given by Eq. \eqref{eq:EnergyOmega}.
In Appendix \ref{app:A}, we obtain $\zeta$ from Kubo's Linear Response Theory~\cite{1966-Kubo-RPP}. Comparing the above expression with Eq. \eqref{app:zetafin}, it is clear that both approaches lead to identical $\zeta$. The negative sign of a transport coefficient results from the presence of $\theta(-t)$ in the response function of the black hole horizon~\cite{2016-Bhattacharya.Shankaranarayanan-PRD}.

\subsection{Fix the constant by mapping to the macroscopic physics}

From Eq.~\eqref{phi_0}, we have:
\begin{equation}
\langle\hat{\varphi}^2_{\omega_H}(0)\rangle =
\frac{1}{2m_*^2}=  \frac{C^2}{2} \, . \label{eqphio2}
\end{equation}
Substituting the above expression in Eq. \eqref{Zeta5-Final}, we get: 
\begin{equation}
\label{Zeta6}
\zeta[\omega_{\rm IR}] = -\frac{1}{8}\frac{\omega_{\rm IR}}{E_\beta(\omega_{\rm IR})} \, .
\end{equation}
In the hydrodynamic limit ($\mathbf{k_*}\rightarrow 0$), Eq.~\eqref{Dispersion} reduces to
\begin{equation}
\omega_{\rm IR} = \frac{1}{C}  
\label{eq:omegaIR}
\end{equation}
Substituting this in Eq.~\eqref{Zeta6} and using Eq. \eqref{eq:EnergyOmega}, we have: 
\begin{equation}
\label{Zeta7}
\zeta[\omega_{\rm IR}] = -\frac{1}{4}  \tanh{\frac{1}{2C}}\, . 
\end{equation}
We can determine the value of $C$ by demanding that the above expression matches with the expression derived by Damour~\cite{1982-Damour-Proc}:
\begin{equation}
\zeta= - \frac{1}{16\pi} \, .
\end{equation}
This leads to the following: 
\begin{equation}
\tanh{\frac{1}{2C}}= \frac{1}{4\pi} \quad \Longrightarrow \quad \frac{1}{C} = \ln\left(\frac{4\pi + 1}{4\pi - 1}\right).
\end{equation}
Thus we have $C \simeq 6.2699$ or $m_* \simeq 0.1595$. 
This implies that the lowest energy excitation in the horizon fluid is approximately $ 0.1595\ T$. Given that $\delta H_{\rm eff}  = -T\delta S$,  the smallest possible change in entropy is $0.32072$, implying that entropy is quantized. It is worth noting that entropy quantization does not occur unless there is a mass gap in the spectrum. 
Within our approach, if the black hole horizon evolves slowly, one can assume the energy spectrum of the system to be dominated by low-energy excitations as in the case of adiabatic quantization of entropy~\cite{1974-Bekenstein-NCL}. 

\section{Thermalization rate from effective field theory}
\label{sec:Thermalization}

In the previous section, we showed that the effective field theory Hamiltonian \eqref{H_Th_Defn} could provide the transport coefficient of the horizon fluid. In this section, we show that the effective theory can be used to predict the thermalization rate of the 
infalling matter-energy to the black hole. 
We do this in two steps. First, we show that the rate of thermalization follows from the horizon-fluid description~\cite{1982-Damour-Proc,1986-Price.Thorne-PRD,1986-Thorne.etal-Membrane}. Second, we use the effective field theory and obtain the scrambling time corresponding to the homogeneous perturbations. 

The phenomenon of rapid delocalization of quantum information in thermal states is referred to as scrambling~\cite{2008-Sekino-Susskind-JHEP}. The	typical	time	scale	of	these	phenomena	is referred to as	scrambling	time~\cite{2008-Sekino-Susskind-JHEP}. For black holes, the scrambling time can be thought of as the time taken by the information content of the infalling matter-energy to be spread into the black hole system. Since most black hole DOFs are expected to exist on its horizon, the scrambling time is defined as the time taken for this information to travel across the entire horizon area or, more precisely, the entire Hilbert space representing the black hole horizon~\cite{2011-Lashkari.etal-JHEP}.

\subsection{Thermalization follows from the horizon-fluid}

Even in weakly interacting systems, the thermalization process is complex. In addition, thermalization is complicated for the event horizon. We consider an analogous model to break this complicated process into simple steps. As we show, using the analogous model makes the discussion more transparent. 

Consider the interaction between a large container of hot gas and a small box of identical gas molecules at a considerably lower temperature. At first, there are only a few gas molecules with relatively low energy. As a result, the energy distribution of the gas molecules deviates from the distribution in thermal equilibrium. However, when the energy of the cold-gas molecules grows, the entire gas relaxes to thermal equilibrium. 
We can approximate this process with two steps: 
\begin{enumerate}
\item After mixing, the entire gas system moves to a state away from the thermal equilibrium. 
\item After some time, the entire gas system thermalizes to a new thermal equilibrium with a lower temperature. 
\end{enumerate}
We repeat this process many times such that  
\begin{enumerate}
\item[A]  Cold gas is brought into contact with hot gas in many uniform steps. At each step, only a minimal amount of cold gas gets mixed into the gas system.
\item[B] This small amount of cold gas is allowed to thermalize with the gas system at each step. Also, let us impose the condition that more cold gas comes in contact \emph{only} with the part of the preceding cold gas thermalized in the last step~\footnote{This is an artificial step and unlikely to occur in terrestrial experiments.}.
 \end{enumerate}
Two physical inputs are required to quantify the process: (i) Since the amount of cold gas is tiny, the thermalization process is stochastic, and (ii) the mean free route of the gas molecules in the gas system does not vary much as a result of mixing. Let $N$ represent the total number of hot gas molecules thermalized, equal to $N= N_0 + \delta N$, where $N_0$ represents the initial number of hot gas molecules, and $\delta N$ represents the number of cold gas molecules thermalized upon mixing. Then the rate of thermalization for this gas system is given by the Langevin equation~\cite{1966-Kubo-RPP}:
 \begin{equation}
 \frac{d^2N}{dt^2} = \Gamma \,  \frac{dN}{dt} + N_{\rm Noise}, \label{LThermalise}
 \end{equation}
where $\Gamma$ is the damping coefficient that depends on the gas properties, and $N_{\rm Noise}$ is stochastic. 

We now apply this physical process for the black hole event horizon.
For a distant observer, the extreme red-shifting of the modes results in the average energy of the external-field modes colliding with the black hole being significantly less than the black hole temperature~\cite{2018-Bhattacharya.Shankaranarayanan-IJMPD}. The quantum fluctuations smear the horizon on an invariant distance of order $\sqrt[3]{M}$~\cite{1997-Casher.etal-NPB,Srivastava:2020cdg}, which acts as a cut-off~\cite{2018-Bhattacharya.Shankaranarayanan-IJMPD}.
Thus, the event horizon is analogous to a big container of hot gas with large entropy ($S_{H}$)~\cite{2018-Bhattacharya.Shankaranarayanan-IJMPD}.  Like the gas system, any matter energy falling into the black hole can be modeled as a perturbation with small entropy, i. e.,  $\Delta S_H \ll S_H$. 
Notably, we have not used any of the event horizon's properties except that of high redshift up to this point.

For an outside observer, the matter falling into the horizon appears to be spread out on the horizon~\cite{1986-Thorne.etal-Membrane}. Any additional matter falling thermalizes due to interaction with this already fallen matter-energy layer. One can formally write the following Hamiltonian of the interaction of the event horizon with external matter fields:
 \begin{equation}
 H_{\rm Total}= H_{\rm Isolated~event-horizon} + H_{\rm External~matter}+ H_{\rm Interaction} \, .
 \label{HTot}
 \end{equation}
The physical content is identical to action \eqref{eq:Formalaction}.
Like in the previous section, we focus on \emph{macroscopic homogeneous perturbations of the event-horizon} and use the effective Hamiltonian \eqref{H_Th_Defn} to describe the evolution of the dynamical horizon. 

The first step towards understanding black hole horizon thermalization is to obtain a rate equation like Langevin equation \eqref{LThermalise}. This is done by first expressing the Raychaudhury equation's homogeneous part in terms of the area expansion coefficient ($\theta_H$):
\begin{equation}
\frac{d\theta_H}{dt} = g_H \, \theta_H 
-\frac{1}{2}{\theta_H}^2 - 8\pi T_{\alpha\beta}\xi^\alpha\xi^\beta. \label{RC}
\end{equation}
where $g_H (= 2 \pi T)$ is the surface gravity on the horizon. Since the above equation is non-linear in $\theta_H$, it is difficult to identify terms that dominate under different physical situations. Rewriting in terms of  the order parameter $\eta \propto \sqrt{\mathcal{A}}$, we have,
\begin{equation}
 \theta_H = \frac{d}{dt}(\ln{\mathcal{A}}) \, ; \frac{d\theta_H}{dt}= \frac{1}{\mathcal{A}}\frac{d^2}{dt^2}(\mathcal{A})-\frac{1}{(\mathcal{A})^2}(\frac{d\mathcal{A}}{dt})^2. \label{ThetaA}
\end{equation}
where $\mathcal{A}$ is the area of the cross-section of the null congruence on the event horizon. Though Eq.~\eqref{RC} is exact, in what follows, we shall also assume $\theta_H$ to be small. Writing $\mathcal{A}= \mathcal{A}_0+ \Delta\mathcal{A}$, where, $\Delta\mathcal{A} \propto \Delta S_H$ is the change in this area over some constant base value $\mathcal{A}_0 \propto S_0$. Thus, the Raychaudhuri equation \eqref{RC} reduces to~\cite{2015-Bhattacharya.Shankaranarayanan-IJMPD}
\begin{equation}
\frac{d^2\Delta S_H}{dt^2} - \gamma \, \frac{d\Delta S_H}{dt} = 
\mathcal{S}_{\rm Noise}. \label{sThermalise}
\end{equation}
where $\gamma = g_H = 2 \pi T$, and ${\cal S}_{\rm Noise}$ is the stochastic term. In Ref. \cite{2015-Bhattacharya.Shankaranarayanan-IJMPD}, the authors derived the Raychaudhuri equation \eqref{RC} from the Langevin equation by retaining quadratic terms.

Eq. \eqref{sThermalise} is a crucial relation, and we stress the following points:  
To begin by comparing Eqs. \eqref{sThermalise} and \eqref{LThermalise}, we can see that the number of degrees of freedom on the event horizon is equivalent to the number of molecules.
 Since the number of degrees of freedom on the event horizon is directly related to black hole entropy, the above equation is the rate equation of thermalization for the perturbed black hole ($\Delta S_H$).
Second, the negative \emph{damping coefficient} implies that this process continues forever due to the one-way nature of the event horizon. This provides another way of understanding the exponential increase of the black hole entropy. 
Third, the equation provides a way to calculate how fast the black hole and the external matter get thermalized. Due to the negative coefficient, thermalization occurs rapidly~\cite{2017-Ropotenko-Arxiv}. 
Lastly, the time required for the external matter-energy information to become inaccessible is $\ln(S_H)/{T}$~\cite{2007-Hayden.Preskill-JHEP,2008-Sekino.Susskind-JHEP,2011-Susskind-Arxiv}. Thus, the horizon fluid provides another way of understanding the scrambling time for black holes.

\subsection{Thermalization rate and bulk-viscosity}

In this subsection, using Eq.~\eqref{sThermalise}, we use the perturbed CFT model of the horizon fluid to relate the thermalization rate to bulk viscosity. To do so within the effective field theory framework, we rewrite the two-step process for the \emph{classical} gas system:
\begin{multline*}
\vert {\rm Initial~equilibrium~state} \rangle_{\rm hot~gas} + 
\vert {\rm Thermal~state} \rangle_{\rm cold~gas} \\ \xrightarrow{\rm Step~1} 
\vert {\rm Out~of~equilibrium} \rangle_{\rm hot~gas}
\xrightarrow{\rm Step~2} \vert {\rm Final~equilibrium~state} \rangle_{\rm hot~gas}
\end{multline*}
The second step, which is our focus, describes the relaxation of \emph{  classical} gas from an out-of-equilibrium distribution to thermal distribution. The thermalization rate can be quantified as follows: On average, the lower the energy of the colder gas molecule compared to the hotter gas temperature, the longer it takes for the molecule to be thermalized. Thus, the thermalization rate is related to the difference in the energy of molecules.

Similarly, the thermalization of external matter-energy falling into the black hole is described as:
\begin{multline*}
\!\!\!\!\! \vert {\rm Initial~ground~state} \rangle_{\rm horizon} 
+ \vert {\rm Excited~state} \rangle_{\rm matter~field} 
\xrightarrow{\rm Step~1} 
\vert {\rm Excited~state} \rangle_{\rm horizon}  \\
+ \vert {\rm Ground~state} \rangle_{\rm matter~field} \xrightarrow{\rm Step~2} \vert {\rm Final~ground~state} \rangle_{\rm horizon} + \vert {\rm Ground~state} \rangle_{\rm matter~field}\, .
\end{multline*}
{  While the overall process of thermalization of the black hole horizon is similar to the classical gas system, there are subtle differences in the final configuration for the two cases. While the gas system is assumed classical, the black hole is intrinsically quantum. In quantum field theoretic language, the zero particle state is denoted by $ \vert {\rm Ground~state} \rangle_{\rm matter~field}\ $.} Classically, the second step describes the dynamical evolution of the perturbed black holes. This evolution is similar to the dynamics of the viscous fluid~\cite{1982-Damour-Proc,1986-Thorne.etal-Membrane, 2016-Bhattacharya.Shankaranarayanan-PRD,2017-Cropp.etal-PRD,2016-Lopez.etal-PRD}. Like in the gas system, the difference in the energy \emph{governs} the thermalization rate.

The effective Hamiltonian \eqref{H_Th_Defn} of the scalar field theory, as opposed to the Langevin dynamics \eqref{sThermalise}, provides two crucial features about the thermalization of the external matter-fields: 
First, \emph{not all modes ($\omega$) get thermalized at the same rate}. The presence of non-zero $m_*$ implies that $\omega \geq m_*$ and  $\omega < m_*$ behave differently. 
Since the exchange of modes, $\omega \geq m_*$ with the event horizon is allowed, these modes thermalize with the event horizon. 
However, this is not the case for $\omega < m_*$ modes. 
Second, Eq. \eqref{HTot} implies that energy conservation holds for all modes. More precisely, creating a low-energy mode of the horizon fluid should be possible via the annihilation of a low-energy mode of the infalling matter field. 
So, what happens to these modes? For the stationary black hole, the event horizon is described by a pure CFT~\cite{1998-Kaul.Majumdar-PLB,1999-Carlip-PRL,2001-Koga-PRD,2001-Hotta.etal-CQG,2002-Hotta-PRD,2010-Barnich.Troessaert-PRL,2011-Carlip-Entropy,2011-Kaul.Majumdar-PRD,2016-Donnay.etal-PRL}, and {all modes} are thermalized. Hence, the low-energy modes get thermalized or scrambled at a much slower rate as the horizon fluid evolves from a perturbed CFT to a pure CFT. 

Now, we present a quantitative estimate of the thermalization rate. Given that the effective scalar field ($\varphi$) is related to the order parameter ($\eta$) for the homogeneous process [cf. \eqref{eq:HF-Heff}], the Langevin equation for the black hole horizon \eqref{sThermalise} can be rewritten as
\begin{equation}
    \frac{d^2\varphi}{dt^2}= \gamma\frac{d \varphi}{dt} + \mbox{noise term} .\label{PhiLamgevin}
\end{equation}
Using the Green-Kubo relation for a thermal bath at temperature $T$, $\gamma$ is given by~\cite{1966-Kubo-RPP}
\begin{equation}
\label{eq:gamma-def}
\gamma \sim \frac{1}{T}\int_0^{\infty} dt 
\langle \partial_t\varphi(t)\partial_t\varphi(0) \rangle \,,
\end{equation}
where $\langle \partial_t\varphi(t)\partial_t\varphi(0) \rangle$ is the two-point auto-correlation function of $\partial_t\varphi$. Since, $\varphi$ 
is proportional to excess-entropy density, $\partial_t \varphi$ 
corresponds to the excess-entropy current density. Physically, the RHS of the above expression corresponds to the correlation of the excess entropy current density of the horizon-fluid~\footnote{Since we only consider homogeneous perturbations, the spatial derivatives vanish.}

From Hamiltonian \eqref{H_Th_Defn}, the equations of motion of $\varphi$ is $\partial_{\mu} \partial^{\mu} \varphi = m_{\rm eff}^2 \varphi$. As mentioned above, at the critical point, $m_{\rm eff} = 0$, leading to 
$\partial_{\mu} \partial^{\mu} \varphi =0$, 
which is the conservation of excess entropy current density $(\partial_{\mu} J_{S}^{\mu} = 0)$. This supports the initial assumption that thermalization stops at the critical point. (For a related discussion on electrical resistivity from current-current correlation, see Ref.~\cite{2012-Hartnoll-Hofman-PRL}.)

Away from the critical point, $\langle \partial_t\varphi(t)\partial_t\varphi(0) \rangle$ is a marginal operator with scaling dimension $\Delta= 3+2\gamma_{\varphi}$~\cite{Peskin:1995ev}, i. e.,
\[
\langle \partial_t\varphi(t)\partial_t\varphi(0) \rangle \sim \frac{1}{\vert t\vert^{\Delta}}
\sim m_{\rm eff}^{\Delta} \, , \]
where $\gamma_{\varphi}$ is a tiny positive number near the critical point and exactly zero at the critical point. The second scaling arises because the only relevant energy scale in the model is $m_{\rm eff}=Tm_*$. As discussed in the previous section, $m_{\rm eff}$ is proportional to the infrared cutoff $\omega_{IR}$ [as shown in Eq. \eqref{eq:omegaIR}]. Substituting the above relation in Eq. \eqref{eq:gamma-def}, we have:
\[
\gamma\sim \frac{m_{\rm eff}^{(2+2\gamma_{\varphi})}}{T} \sim 
T^{1 + 2\gamma_{\varphi}} \, .
\]
It is worth noting that, in the earlier subsection, we obtained a similar expression for $\gamma$ ($\gamma \sim T$) when we identified the Langevin equation with the Raychaudhuri equation \eqref{sThermalise}.

Let us now look at thermalization through 
dissipation via bulk viscosity. At the critical point $m_{\rm eff} \to 0$ implying $C \to \infty$. From Eq.~\eqref{Zeta7}, we see that close to the critical point, $\zeta \sim 1/(2 C)$. Thus, both $\gamma$ and $\zeta$ flow to zero as the system reaches the critical point. {Thus, the coefficient of bulk viscosity determines the scrambling time.}

Our analysis reveals an intriguing fact: {mass gap is critical in determining the bulk viscosity ($\zeta$) and damping coefficient} ($\gamma$), both of which affect the thermalization rate. For macroscopic black holes, the thermalization process described by the classical equation \eqref{sThermalise} is a good approximation as most of the energy falling into the black hole has been thermalized. However, as the system approaches the critical point, $m_{\rm eff}$ decreases, and the equation governing thermalization deviates more from the classical equation \eqref{sThermalise}. The mass gap $m_{\rm eff}$ (or ${1}/{C}$) is a measure of the deviation of the near-horizon geometry from that of the asymptotically flat, stationary black hole space-time. For smaller-size black holes, $m_{\rm eff}$ is higher, and more of these modes thermalize slowly, which may hold the key to how information escapes from a black hole at late times. This is being investigated. Finally, we would like to emphasize that our arguments are model-independent to a large extent. This is because theories near the critical point are universal. Thus, the argument presented here is valid if we assume that the theory is approaching a critical point.

\section{Microscopic toy model corresponding to effective field theory}
\label{sec:Toymodel}

The last two sections demonstrated that the effective field theory corresponding to homogeneous horizon perturbations can account for the bulk viscosity of the horizon fluid, which in turn can explain scrambling time. 
This section proposes a microscopic toy model for the effective Hamiltonian \eqref{H_Th_Defn}. 
Like the construction of effective Hamiltonian, the microscopic toy model considers \emph{two different aspects of the black hole horizon}.   First, the model must incorporate near-horizon symmetries of the stationary black hole~\cite{1998-Kaul.Majumdar-PLB,2011-Kaul.Majumdar-PRD,2001-Koga-PRD,2001-Hotta.etal-CQG,2002-Hotta-PRD,1999-Carlip-PRL,2011-Carlip-Entropy}, which we have already discussed in some detail in the context of constructing an effective field theory. 
Second, the model must incorporate the physics of transport phenomena of horizon-fluid~\cite{2016-Bhattacharya.Shankaranarayanan-PRD,2017-Cropp.etal-PRD}. In this work, this will mean only the phenomenon of bulk viscosity of the horizon fluid. 
However, these two aspects of the black hole horizon do not necessarily constrain the microscopic model, as many microscopic models can satisfy both of these aspects. 
In this work, we consider {\sl Eight-vertex Baxter model}~\cite{1971-Baxter-PRL,1973-Baxter.Wu-PRL,1976-Baxter-JStatPhys,1977-Baxter-JStatPhys,1978-Baxter-JStatPhys,2016-Baxter-Book} as an illustration of the above mentioned ideas. We demonstrate that this integrable model satisfies both horizon requirements.  

The model possesses the following properties, which serve as crucial components in the microscopic model building of the horizon-fluid: To begin with, it has a {\sl lattice} Virasoro algebra that corresponds ${\mathcal S}1$ diffeomorphism symmetry~\cite{1971-Kadanoff.Wegner-PRB,1987-Itoyama.Thacker-PRL}. 
 Second, it comprises two staggered 2D Ising lattices and has the same free energy density as the 2D Ising model. However, the symmetry of the two-sublattice model is quite different from that of the conventional Ising model. As a result, the critical indices of the Baxter solution are generally different from those of Ising~\cite{1971-Kadanoff.Wegner-PRB}. Third, it exhibits a second-order phase transition. In the continuum limit, it is an Integrable Field Theory near the critical point and is a CFT at the critical point~\cite{1986-Thacker-Physica,1987-Itoyama.Thacker-PRL,1989-Itoyama.Thacker-NPB}.

To demonstrate that the Eight-Vertex Baxter model can indeed be used as a microscopic toy model, we follow three steps:
\begin{enumerate}
    \item Adopt the Baxter model for the black hole horizon. This is done by the projection of two planes onto the surface of a sphere. 
    \item Show that the eight-vertex Baxter model incorporates $l_n^{\rm Diff}$ diffeomorphism symmetry. 
    \item Show that in the continuum limit, the microscopic toy model leads to effective Hamiltonian \eqref{H_Th_Defn}.
\end{enumerate}

\subsection{Adopting Baxter Model for black hole horizon}

\begin{figure}[!hbt]
    \centering
    \includegraphics[width=0.7\linewidth]{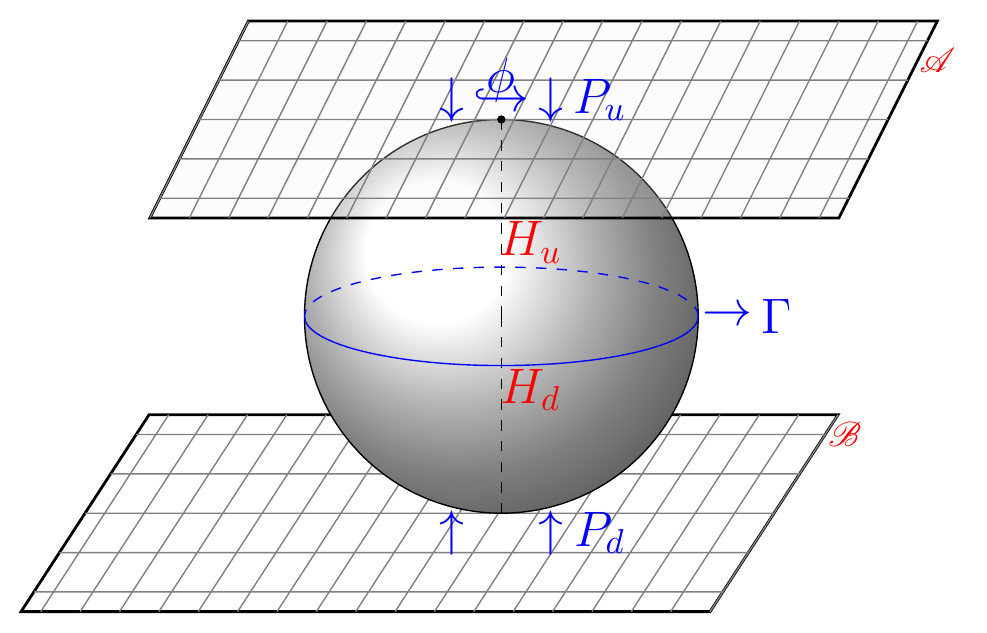}
  \caption{The projection of the 8-vertex Baxter model from the two sub-lattices to $S^2$ surface of the horizon.} 
\label{fig:Stereographic}  
   \end{figure}
The alert reader may wonder how the Baxter model can be adapted to the cross-section of a black hole event horizon, a $S^2$ surface. As shown in \ref{fig:Stereographic}, this model can be adopted on the two hemispheres of the $S^2$ surface through projection from two Baxter lattices. Let $P_u$ ($P_d$) denote the map corresponding to the projection $\mathscr{A}\rightarrow H_u$ ($\mathscr{B}\rightarrow H_d$). For the consistency of the model, we need to impose the condition $P_u^{-1}\circ\{\Gamma\} \equiv P_d^{-1}\circ\{\Gamma\}$, where $P^{-1}$ denotes the inverse map, and $\Gamma$ is the equatorial plane of the $S^2$ surface. 

The above condition retains the periodic boundary condition of the Baxter model. The projection allows relating the Euclidean boost parameter of the Baxter model~\cite{1989-Itoyama.Thacker-NPB} to the azimuthal angle in the spherical polar coordinate. We can relate the Virasoro algebra (corresponding to the $\mathcal{S}1$ diffeomorphism) of the Euclidean boost parameter to the $\mathcal{S}1$ diffeomorphism of the azimuthal angle in the horizon-fluid model. Thus, the projection retains the model's main physical features~\cite{1989-Itoyama.Thacker-NPB} and incorporates a representation of the $\mathcal{S}1$ diffeosymmetry in the model. It is the diffeosymmetry of the azimuthal angle for the black hole system. Like any other lattice approach, the challenge consists in showing that a continuum limit exists for which the effective action has diffeomorphism invariance~\cite{2009-Hamber-GRG,2012-Wetterich-PRD}. As we show, this model reduces to effective Hamiltonian~\eqref{H_Th_Defn}, which preserves the symmetry. Our next task explicitly shows that the Eight-vertex Baxter model possesses the Virasoro algebra. 

\subsection{Eight-Vertex model and deformed CFT}
\label{sec:3}
 
The model has eight possible arrangements of arrows at a vertex with four distinct Boltzmann weights $a, b, c, d$. These satisfy two constraints~\cite{2016-Baxter-Book}: 
\begin{equation}
\label{eq:weightsdef}
\frac{c d}{a b} = \frac{1 - \Gamma}{1 + \Gamma}~;~ 
\frac{a^2 + b^2 - c^2 - d^2}{2 (a b + c d)} = \Delta
\end{equation}
For constant $\Gamma$ and $\Delta$, there exists a one-parameter family of Boltzmann weights ($w$) which satisfy the star–triangle relations and, hence, the eight-vertex model has a one-parameter family of commuting transfer matrices~\cite{2016-Baxter-Book}. This allows one to parameterize the Boltzmann weights explicitly in terms of {\it spectral variable} ($u$):
\begin{equation}
\begin{array}{ll}{a=\operatorname{snh}(\lambda-u)} 
& {b=\operatorname{snh} u}\\ 
{c= k \operatorname{snh} \lambda,} & {d=k \operatorname{snh}(\lambda-u)}\end{array}
\label{elliptic}
\end{equation}
where, $k$ is the elliptic modulus, and $\operatorname{snh} \ u = -i\ \operatorname{sn}(iu)$.  It has been shown that the transfer matrix of the eight-vertex model commutes with the ${\rm XYZ}$ Hamiltonian~\cite{1970-Sutherland-JMP}:
\begin{equation}
{H}_{\rm XYZ} = -\frac{1}{2} \sum_{j = 1}^N J_{\sigma} \sigma^{\sigma}_j\sigma^{\sigma}_{j+1} 
\quad \mbox{where}  \quad \sigma = x, y, z \, .
\label{XYZ}
\end{equation}
The coupling constants are related to the weights by the relation: 
%
 $J_x:J_y:J_z 
 = 1: \Gamma : \Delta $. 
%
The spins $\sigma_n$'s are related to the vertex weights by the vertex operator ($V_n$):
{\small
$$
 V_n= \frac{1}{2} \left[a+c+ [a-c] \sigma_n^z\sigma_{n+1}^z + 
 [b+d]  \sigma_n^x\sigma_{n+1}^x+ \sigma_n^y\sigma_{n+1}^y\right] 
 $$
}

The row-to-row transfer matrix can be expressed as 
\begin{equation}
 T(u)= \lim\limits_{N \to \infty} 
 {\rm Tr}\left(V_{-N} \, V_{-N+1} \cdots V_N \right)  \, .     
 \label{RTM}
\end{equation}
The transfer matrix can be expanded formally around a point $u  = u_0$, as
\begin{equation}
\!\!\! \ln T(u) = \sum_{n = 0}^{\infty} I_n (u - u_0)^n~;~ 
I_{n} := \left.\frac{1}{n !} \frac{d^{n} \ln \mathbf{T}(u)}{d u^{n}} \right|_{u=u_{0}} 
\label{LnT}
\end{equation}
$I_n$ ($n \geq 1$) can be interpreted 
as the operators that couple with $(n + 1)$ neighboring sites~\cite{2016-Baxter-Book}. $I_1$ corresponds to the Hamiltonian of a spin chain with the nearest neighbor (\ref{XYZ}):
\begin{equation}
I_1 = - H_{\rm XYZ} \equiv 
\sum_{j = -\infty}^{\infty}\mathcal{H}_{XYZ}(j,j+1).
\end{equation}

It is more convenient for our purpose to describe the model in terms of the Corner Transfer Matrix. The corner transfer matrix operator can be viewed physically as connecting semi-infinite rows of arrows with a semi-infinite column of arrows of one quadrant of the lattice. In the thermodynamic limit, the following relation holds 
\cite{2016-Baxter-Book}:
\begin{equation}
 \mathcal{A}(u)= \exp\left[-\frac{\pi u}{2K} \, \mathcal{L}_0\right] \, , \label{Full A}
\end{equation}
where $K$ is a complete elliptic integral associated with modulus $k$ and 
\begin{equation}
\mathcal{L}_0 = \frac{2K}{\pi}\sum_{j = -\infty}^{\infty}j\mathcal{H}_{XYZ}(j,j+1). \label{L_0}
\end{equation}
To keep the calculations transparent, we set $\Gamma = 0$, i. e., $cd = a b$ in \eqref{eq:weightsdef}. This corresponds to the condition $J_z  = 0$ in the Hamiltonian (\ref{XYZ}), which is the well-known XY model~\cite{2016-Baxter-Book}. ${\cal L}_0$ in Eq. (\ref{L_0}) is diagonalized by the operators:
\begin{equation}
{\Psi(l)=N_{l} \int d \alpha \, e^{-\imath \alpha l \pi / 2 K} \chi(\alpha)} 
\label{eq:Psidef}
\end{equation} 
where $N_{l}$ is the normalization constant, and
%
${\chi(\alpha)=\operatorname{sn} \alpha \, a^{\alpha}(\alpha)+\imath \sqrt{k} \operatorname{cn} \alpha \, a^{y}(\alpha)} \, .$
%
The integration over $\alpha$ is over one complete real period of the 
elliptic functions from $-2 K + \imath K'/2$ to $2 K + \imath K'/2$. 
Itoyama and Thacker~\cite{1986-Thacker-Physica,1987-Itoyama.Thacker-PRL,1989-Itoyama.Thacker-NPB} showed that ${\cal L}_0$ could be expressed:
\begin{equation}
\mathcal{L}_0 = \sum\limits_l l : \bar{\Psi}(l)\Psi(l) : + \,  h \, ,
\end{equation}
where $h$ is a constant and $:~:$ refers to as normal-ordered product. ${\cal L}_0$ is embedded into a Virasoro algebra as a central
element~\cite{1986-Thacker-Physica,1987-Itoyama.Thacker-PRL,1989-Itoyama.Thacker-NPB}.   The normal ordering is defined by the relations,
$\Psi(l)|h\rangle = 0 (\forall \ l\geq 1),  
\overline{\Psi}(l)|h\rangle = 0, (\forall \ l\leq 1) \, .$ 
%
Other Virasoro operators ${\cal L}_n$ 
can be constructed from these momentum space operators~\cite{1989-Itoyama.Thacker-NPB}.
From \eqref{eq:Psidef}, it follows that,
\begin{equation}
[\mathcal{L}_n,\mathcal{L}_m] = (n-m)\mathcal{L}_{n+m} + \frac{1}{12}c(n^3-n)\delta_{n+m,0}.  \label{L_Virasaro}
\end{equation}
As noted in Ref. \cite{1989-Itoyama.Thacker-NPB}, the physical Hilbert space built from the state $|h\rangle$ forms the highest weight representation of the Virasoro algebra. Since the eigenvalues of $\mathcal{L}_0$ are doubled due to the zero modes of the operator $\Psi(0)$ and $\overline{\Psi}(0)$, the highest weight vector forms a two-dimensional representation under parity conjugation. At the critical point, the central charge $c= 1$ and $h={1}/{8}$.

Using the following classical generators ($l_n^{Diff}$),
\begin{equation}
l_n^{Diff}= -\frac{1}{2}\zeta^{n+1}\frac{d}{d\zeta}- \frac{1}{2}\frac{d}{d\zeta}\zeta^{n+1} \, ,\label{Diffeo_l_n_Defn}
\end{equation}
we can obtain other Virasoro algebras different from the one described above. The difference is that $l_n^{\rm Diff}$ are generators of diffeo-transformation of the spectral rapidity parameter or the Euclidean boost parameter ($\alpha$)~\cite{1985-Gervais-PLB,1989-Itoyama.Thacker-NPB}. The corresponding Virasoro algebra can then be constructed by defining the following $\mathcal{L}_n^{\rm Diff}$:
{\small
\begin{eqnarray}
\mathcal{L}_n^{diff} = :\int_{-K}^{3K}\frac{d\beta}{2\pi}B(\beta+2K-\imath K')l_nB(\beta):+ h\delta_{n,0} 
= \sum\limits_l(l+\frac{n-1}{2}):\bar{\psi}(l)\psi(l+n):+ h \delta_{n,0}. \label{L_N_Diffeo}
\end{eqnarray} 
}
This demonstrates that the eight-vertex model possesses the Virasoro algebra given by \eqref{L_N_Diffeo}, which holds the key to incorporating near-horizon $S1$ diffeo symmetry in the model of the horizon-fluid. The 2-D Euclideanized space-time $(\tau,q)$ can be identified with the 2-D  Euclidean space $(x_1,x_2)$ on which the horizon-fluid resides. The rapidity or the boost parameter in a Euclideanized space-time {corresponds to the} rotation angle. A closer look reveals that, in this case, the rapidity is the azimuthal angle depicted in  \ref{fig:Stereographic}. Thus, $l_n^{\rm Diff}$ diffeomorphism algebra of the spectral rapidity corresponds to the $l_n^{\rm Diff}$ diffeomorphism symmetry on the black hole horizon. Thus we see that the microscopic modelling of the horizon-fluid with a mass gap incorporates $l_n^{\rm Diff}$ diffeomorphism symmetry on the black hole horizon. Our next task is to relate the Eight-vertex Baxter model with the effective field theory Hamiltonian \eqref{H_Th_Defn}.

\subsection{Continuum limit and effective field theory} 

Long-range effects dominate the critical properties of the model; hence, a continuum approximation will suffice. The eight-vertex model's continuum limit is a theory of massive Dirac Fermions $(\Psi_1, \Psi_2)$~\cite{1989-Itoyama.Thacker-NPB} 
{
\begin{equation}
\mathcal{S}_{\rm Dirac} =  \int d\tau dq \, \left[\frac{1}{2}i\overline{\Psi}_1\Big(\overleftrightarrow{\frac{\partial}{\partial \tau}}+\overleftrightarrow{\frac{\partial}{\partial q}}\Big)\Psi_1 
+ \frac{1}{2}i\overline{\Psi}_2\Big(\overleftrightarrow{\frac{\partial}{\partial \tau}}-\overleftrightarrow{\frac{\partial}{\partial q}}\Big)\Psi_2 - m(\bar{\Psi}_1\Psi_2+\bar{\Psi}_2\Psi_1) \right]. \label{LagrangianD}
\end{equation}
}
The above action possesses an infinite sequence of conserved densities.
Physically, this implies that besides the total angular momentum, the entire momentum distribution is conserved~\cite{1989-Itoyama.Thacker-NPB}. Interestingly, it turns out that all of these operators can also be written as integrals of local densities in coordinate space,
\begin{equation}
{\cal L}_{n}=\int dq \,  J_{0}^{(n)}(q) \, , 
\end{equation}
where $J_0$ is the zeroth component of a conserved current. $J^{(-1)}_0$ is the Hamiltonian plus the momentum operator, and $J_0^{(0)}$ (at $\tau = 0$) is the first moment of the Dirac Hamiltonian~\cite{1989-Itoyama.Thacker-NPB}. Integrability ensures the operators are related to the infinite sequence of conserved charges, with one for each new ${\cal L}_n$~\cite{1989-Itoyama.Thacker-NPB}. This satisfies another critical requirement of the microscopic model:~an integrable field theory with an infinite number of conserved charges corresponding to an infinite number of symmetries~\cite{1989-Zamolodchikov-Proc}. 

While the action \eqref{LagrangianD} is useful to identify the infinite number of conserved charges, it is rather cumbersome doing the hydrodynamic or long-wavelength limit calculations starting from the action \eqref{LagrangianD}. Hence,  in what follows, we will not be directly using the action \eqref{LagrangianD}. Instead, to make contact with the phenomenological analysis, we turn to the fact that the Free-energy density of the Baxter model is the same as that of a classical 2- D Ising model near the critical point~\cite{2016-Baxter-Book}. On the other hand, the theory of the 2-dimensional Ising model can be described by a theory of a free massive scalar field $\varphi$ in a two-dimensional Euclidean spacetime~\cite{1987-Polyakov-Book}. This can be viewed as a mean-field description of the Baxter model. 

Thus, the 2-D mean-field theory Hamiltonian of the microscopic model can be extended to the following $2+1$ dimensional space-time Hamiltonian:
\begin{equation}
H_{\rm eff }(\varphi) =   \int \bigg[\frac{1}{2}\big(\frac{\partial \varphi}{\partial t}\big)^2+ \frac{1}{2}\big(\nabla\varphi\big)^2 +\frac{m_*^2}{2}\varphi^2\bigg] dA, \label{H_Th_Defn2}
\end{equation}
which is identical to the effective Hamiltonian \eqref{H_Th_Defn}. The above Hamiltonian satisfies the essential requirement of possessing $Z2$ symmetry and 
can be viewed as a mean-field description of the 
horizon-fluid near a critical point. While this mean-field description does not capture all of the details of the quantum states or reproduce the correct scaling exponents, it does describe the horizon-fluid properties, as confirmed in Secs. \eqref{sec:BulkViscosity} and \eqref{sec:Thermalization}.  

\section{Discussion} 
\label{sec:conc}

This work outlines an approach to studying the low-energy physics of the dynamical black hole horizons by constructing an effective field theory. Our starting point is that CFT is a plausible candidate for the effective theory on the horizon. Using the fact that the dynamical (non-stationary) black hole can be viewed as interacting with external fields leads to the condition that the theory is not conformal but must incorporate the ${\cal S}(1)$ diffeomorphism symmetry. Thus, in our approach, the perturbed black holes are described by deformed CFTs, and the deformation scale sets the interaction between the horizon and the external fields. By relating the effective scalar field Hamiltonian \eqref{H_Th_Defn} to the energy density of the horizon fluid, we obtained a relation between the scalar field ($\varphi$) with the order parameter. This enabled us to calculate the bulk viscosity coefficient ($\zeta $) of the horizon-fluid from the correlation functions of the effective field theory Hamiltonian \eqref{H_Th_Defn}. Additionally, the infrared cutoff corresponding to a mass gap in the theory enables a straightforward derivation of $\zeta$ and area quantization. We also constructed a minimal microscopic toy model for the horizon-fluid that reduces to the effective field theory Hamiltonian \eqref{H_Th_Defn}. This model illustrates the construction of a microscopic model of a dynamical horizon.

Our approach allows us to connect two important constants --- the bulk viscosity coefficient $\zeta $ of the horizon-fluid and Bekenstein's quantum of entropy ~\cite{1974-Bekenstein-NCL}. Furthermore, we 
showed that the bulk viscosity coefficient of the horizon fluid determines the time required for black holes to scramble. As thermalization progresses, the perturbed black hole approaches the stationary state, implying that the theory also flows toward the critical point. Thus, we showed that one could use scaling relations for the relevant transport coefficients to observe how the thermalization rate decreases with time, implying a departure from general relativity.

The effective field theory considered here works for bulk viscosity only and cannot describe all the involved processes near the black hole horizon. This is because the effective scalar degree of freedom we have identified can not describe all the processes. However, it demonstrates how such a simple theory can incorporate near-horizon symmetries and account for the bulk viscosity of the fluid. However, the key ideas used in constructing this model are quite generic. They can play an important role in the future in building more realistic effective-theory models for perturbed black holes in asymptotically flat spacetimes. While this has been done with AdS background black holes using the Ads/CFT correspondence, the problem is still unsolved for other types of black holes.

This work assumes CFT can incorporate ${\cal S}(1)$ diffeomorphism symmetry of a stationary black hole; hence a theory describing the event horizon of a stationary black hole should be a CFT. It is important to note that there is no consensus in the literature yet. This is because all the approaches suffer from incompleteness, as it has not been possible to explore the full phase space of the problem. However, a more systematic exploration of the phase space is performed in some special cases~\cite{Averin_2019}, and the results appear to confirm that the theory on the horizon is CFT. Unfortunately, the scope of this work does not permit us to enter the debates on this issue. {We remind the reader here once more, that ideally a full quantum gravity theory is expected to exhibit full diffeo-invariance. However, as we are here only considering an effective field theory on the fixed background of an event horizon spacetime, that constraint does not apply here. [For a more detailed discussion, see Refs.~\cite{2022-Tessarotto1,2022-Tessarotto2}].}

In this work, we have focused on non-extremal black holes. Unfortunately, our approach is insufficient for extremal black holes to arrive at a value for $\zeta$. This is because the assumption that only the stationary black hole is a critical point and the perturbed black hole eventually relaxes to the stationary black hole is not applicable for extremal ones. However, it is instructive to keep in mind that $\zeta$ is independent of $T$ as calculated within our model, so the results from our model do not contradict the known characteristics of an extremal black hole in asymptotically flat spacetime. It is only $\gamma$, as defined by us, that is proportional to $T$ and hence would be zero for extremal black holes. However, this does not raise any problem of consistency.

In this work, we have only considered homogeneous perturbations of the stationary black hole. Therefore, we must develop an effective field theory encompassing general perturbations capable of describing the horizon fluid. Naturally, this complete theory would be richer than the simple scalar field theory discussed here. As a result, we anticipate a more comprehensive and effective theory of the horizon fluid that can adequately describe a system approaching the critical point. Moreover, it should have a representation of the near-horizon symmetry algebra and a known mass gap.

\acknowledgments{
The authors thank M. Bojowald, S. Braunstein, Saurya Das, N. Dadhich, R. K. Kaul, A. Laddha, Late T. Padmanabhan, Subodh Shenoy, and A. Virmani for the discussions. SB thanks Physics Department, IIT Bombay, for its hospitality. We thank the anonymous referees for their critical comments, which helped us in improving the quality of the work. SS is partially supported by the SERB-MATRICS grant.}

\appendixtitles{yes} 
\appendixstart
\appendix
\section{Coefficient of bulk viscosity from linear response theory}
\label{app:A}

In this Appendix, we follow Kubo~\cite{1966-Kubo-RPP} 
and derive the bulk viscosity coefficient of the horizon-fluid from the Hamiltonian~\eqref{H_Th_Defn}. Instead of using 
correlations of the stress-energy tensor of the scalar field theory~\cite{1995-Jeon-PRD}, we compute the bulk viscosity of the horizon-fluid from the autocorrelation function of the current~\cite{1966-Kubo-RPP}. 

To accomplish this, we need to include an interaction Hamiltonian to the microscopic Hamiltonian \eqref{H_Th_Defn}. 
Here, we take a more direct approach. Rather than express the interaction Hamiltonian explicitly in terms of external matter fields, we will describe the external influence on the black hole horizon by examining the black hole horizon's excitation by the infalling matter energy. Recall that the infalling energy-matter increases entropy by an amount, $\delta S$. Thus, an external influence excites the horizon from its ground state. The response is expected to be linear for small perturbations. The change, in this case, is a strain with a non-zero area. From \eqref{eq:phi-tildephi-relation}, it follows that:
\begin{equation}
\langle\tilde{\varphi}\rangle = \frac{1}{2}C\frac{\delta\cal A}{\cal A}. \label{PhiStrain}
\end{equation}
For bulk viscosity processes in the horizon-fluid, the area strain is
\[
\frac{\delta\cal A}{\cal A} = \frac{2\tilde{\varphi}}{C} \, .
\]
Thus to describe bulk viscosity, we  can approximate 
$H_{\rm ext}$ as~\cite{1966-Kubo-RPP}:
\begin{equation}
H_{\rm ext} = - \frac{T}{4}\int K(t) \frac{\delta\cal A}{\cal A} dA =
- T\int K(t) \, \frac{ \tilde{\varphi}}{2C}  \, dA \, . \label{H_ext}
\end{equation}
Thus, the total Hamiltonian of the Horizon to describe the bulk viscosity processes in the horizon-fluid is:
\begin{eqnarray}
H_{\rm HF} = H_{\rm eff}(\varphi)+ H_{\rm ext} =
 \int \bigg[\frac{T}{2}\big(\frac{\partial \tilde{\varphi}}{\partial t}\big)^2+ T \big(\nabla\tilde{\varphi}\big)^2 +\frac{T}{2} m_*^2\tilde{\varphi}^2
- T\frac{1}{2C}\tilde{K}(t)\tilde{\varphi} \bigg]  
dA \, . \label{HTotal01}
\end{eqnarray}
Again performing a rescaling as before, $H_{\rm ext}$ can be expressed as $H_{\rm ext} = \frac{1}{2C}\int K(t)\varphi dA$, where, $K(t)= \sqrt{T}\tilde{K}(t)$. Thus the total Hamiltonian can also be expressed in the form, 
\begin{equation}
H_{\rm HF} =  \int \bigg[\frac{1}{2}\big(\frac{\partial \varphi}{\partial t}\big)^2+  \big(\nabla\varphi\big)^2 +\frac{1}{2 C^2}\varphi^2
- \frac{1}{2C}K(t)\varphi \bigg]  
dA \, \label{eq:HamilHF01}
\end{equation}
The external influence results in a change in $\varphi$, given by 
\[
\Delta\varphi = \varphi_{\rm Excited}- \varphi_{\rm Ground} \, . 
\]
The entropy is maximum if the ground state is taken to be the state of equilibrium of the system; hence, $\delta S$ goes to zero in that state. This means we can set 
$\varphi_{\rm  Ground}= 0$ and $\Delta\varphi = \varphi$.  The following relation gives the response of the fluid:
\begin{equation}
\frac{2\varphi}{C} = \int_{-\infty}^t dt' \int K(t')\varphi(t-t')dA, \quad 
\varphi(t - t') = \langle [\frac{\hat{\varphi}(t)}{2C}, \frac{2\hat{\varphi}(t')}{C}] \rangle
\label{Response}
\end{equation}
where $\varphi(t - t')$ is the response function for the process describing bulk viscosity of the Horizon-fluid, and $\langle\rangle$ denotes the statistical average of the physical variable. Taking a Fourier transform of \eqref{Response} with respect to time, we get:
\begin{equation}
\frac{2\varphi(0)}{C} = \frac{\delta\cal A}{\cal A} = - \zeta[\omega]\tilde{K}[\omega].
\end{equation}

To evaluate the response function ($\varphi(t - t')$) for the process, we proceed by rewriting the field $\hat{\varphi}_*$ (also $\hat{\varphi}$) in terms of the creation and annihilation operators, i. e.,
\begin{equation}
\hat{\tilde{\varphi}}(t,\mathbf{x})= \sum_{\mathbf{k_*}}[\hat{a}_{\mathbf{k_*}}u_{\mathbf{k_*}}(t,\mathbf{x})+ \hat{a}_{\mathbf{k_*}}^{\dagger}u_{\mathbf{k_*}}^*(t,\mathbf{x})],
\label{eq:modeexpansionA}
\end{equation}
where, 
\begin{equation}
u_{\mathbf{k_*}}(t,\mathbf{x}) = \frac{1}{\sqrt{2\pi}}\frac{1}{\sqrt{2\omega_*}} e^{i(\mathbf{k_*}.\mathbf{x}-\omega_* t)}
\end{equation}
and $\mathbf{k_*}$, $\omega_*$, $\mathbf{x}$ and $t$ are dimensionless variables in the above equation. Note that $l_P=1$ in natural units. For the ease 
of notations, we denote the dimensionless space-time coordinates by the same notation, $\mathbf{x}$ and $t$. In the hydrodynamic limit, the corresponding infrared cutoff is much smaller than the dimensions of the total volume (actually area in this case). Thus, we neglect the effect of the extrinsic curvature of the cross-section of the black hole. Hence, in our analysis of the horizon fluid corresponding to a macroscopic black hole, we use plane wave modes; the field theory calculation performed here can be thought of as 'local' compared to the entire area of the horizon cross-section.

The dimensionless part of the Hamiltonian, $\hat{H}_{\rm eff}$ as given by \eqref{HTotal01}, can then be expressed in the frequency space as 
\begin{equation}
\int \bigg[\frac{1}{2}\big(\frac{\partial \tilde{\varphi}}{\partial t}\big)^2+  \big(\nabla\tilde{\varphi}\big)^2 +\frac{1}{2}m_*^2\tilde{\varphi}^2\bigg]dA = \sum_{\mathbf{k_*}}(\hat{a}_{\mathbf{k_*}}^{\dagger}\hat{a}_{\mathbf{k_*}}+\frac{1}{2})\omega_* ,
\end{equation}
which leads to 
\begin{equation}
H_{\rm eff}(\varphi)= T \sum_{\mathbf{k_*}}(\hat{a}_{\mathbf{k_*}}^{\dagger}\hat{a}_{\mathbf{k_*}}+\frac{1}{2})\omega_* , 
\end{equation}
Note that we use the dimensionless variables for frequencies and wavenumbers as we consider the dimensionless part of the Hamiltonian. 
The dispersion relation also follows, given by, 
\begin{equation}
\omega_*^2 = \mathbf{k_*}^2+ {1}/{C^2} . 
\label{DispersionA}
\end{equation}
Of course, an equivalent relation concerning $\mathbf{k}$ and $\omega$ can be written from \eqref{Dispersion}, 
\begin{equation}
T^2\omega_*^2 = T^2\mathbf{k_*}^2+ {T^2}/{C^2} . 
\end{equation}

Following Kubo~\cite{1966-Kubo-RPP}, one can then write down the expression for $\zeta[\omega_*]$ as 
\begin{equation}
\zeta[\omega_{\rm IR}] = \operatorname{Im} \left[\int_0^{\infty}\langle[\frac{\hat{\varphi}(0)}{2C}, \frac{2\hat{\varphi}(t)}{C}] \rangle \, 
e^{-i\omega_{\rm IR} t} dt \right] \, ,
\label{ZetaA}
\end{equation}
where $T\omega_{\rm IR}$ is the energy quanta corresponding to the frequency $\omega_{\rm IR}$ of the field $\varphi$. (As mentioned earlier, $\omega_{\rm IR}$ is dimensionless!) 

From this, we get the following:
\begin{equation}
\label{app:zetafin}
\zeta[\omega_{\rm IR}] = \operatorname{Im} \left[\frac{1}{C^2}\int_0^{\infty}\langle[\hat{\varphi}(0), \hat{\varphi}(t)] \rangle \, 
e^{-i\omega_{\rm IR} t} dt \right] \, ,
\end{equation}
In the hydrodynamic limit, this gives the same expression for $\zeta$ that we obtained in section \eqref{sec:BulkViscosity} from \eqref{ZetaF} and \eqref{deltaphiOp}. The only difference is the theta function, which one can incorporate in the response $\varphi(t)$. Essentially, it refers to expressing it in terms of the scalar field's ($\varphi$) thermal correlation function/thermal Green's function. From this point, we can determine the value of $\zeta$ as in Sec. \eqref{sec:BulkViscosity}. 

\section{Teleological boundary condition}
\label{app:Teleo}

To compute the correlation function 
$\rho_{\delta \bar{P} \delta \bar{P}}$, it is necessary to understand how the horizon fluid reacts to the external environment. Remarkably, the event horizon's response to any external impact is anti-causal. Specifically, if matter energy falls over the event horizon, the event horizon expands until the matter energy goes through the horizon. As the event horizon of a black hole is defined globally in the presence of the future light-like infinity, this is not unphysical~\cite{1986-Thorne.etal-Membrane}.

Due to this peculiar property of the horizon, the horizon fluid exhibits an anti-causal response, meaning that the response of the horizon occurs before the external influence~\cite{1986-Thorne.etal-Membrane}. From a fluid's perspective, the system is initially out of equilibrium and evolves slowly towards equilibrium; external influence brings the system to equilibrium, preventing any further evolution from that state. This is referred to as the teleological nature of the horizon~\cite{1986-Thorne.etal-Membrane}. It has been demonstrated in the literature that if the system exhibits an anti-causal transport process, then the anti-causal transport coefficients have the opposite sign of their causal counterparts~\cite{1996-Evans.Searles-PRE}. For normal fluids, external influence disrupts the equilibrium. For horizon fluid, the opposite is true; the system tends toward equilibrium in anticipation of external influences, such as the infusion of energy into the fluid.

\section{Explicit evaluation of bulk viscosity}
\label{App:QFTBulkViscosity}

To obtain $\zeta$,  we need to evaluate $\big[\delta\hat{\varphi}(t),\delta\hat{\varphi}(0)\big]$. To do this, we rewrite $\hat{\delta\varphi}$ in terms of $\hat{\varphi}$, i. e., 
\begin{equation}
\hat{\delta\varphi}= \hat{\varphi}- \varphi_{\rm S} \hat{\mathbf{1}}  \Longrightarrow 
\big[\delta\hat{\varphi}(t),\delta\hat{\varphi}(0)\big]= \big[\hat{\varphi}(t),\hat{\varphi}(0)\big]
\label{deltaphiOp}
\end{equation}
To obtain the above commutation relation of the field operator, we write the Hamiltonian corresponding to the Lagrangian \eqref{Ldensity}: 
\begin{equation}
H_{\rm HF} =  \frac{1}{2} \int \bigg[\big(\frac{\partial \varphi}{\partial t}\big)^2+  \big(\nabla\varphi\big)^2 +\frac{1}{C^2}\varphi^2\bigg]  
dA \, \label{eq:HamilHF}
\end{equation}
As we will see, it is easier to analyze rescaled variables  $\varphi= \sqrt{T}\tilde{\varphi}$, (where $T$ is a constant). Thus, the above Hamiltonian becomes: 
\begin{eqnarray}
 H_{\rm HF} = 
\frac{T}{2} \int \bigg[\big(\frac{\partial \tilde{\varphi}}{\partial t}\big)^2+ \big(\nabla\tilde{\varphi}\big)^2 + m_*^2\tilde{\varphi}^2
 \bigg]  
dA \, . \label{HTotal}
\end{eqnarray}
Now we proceed by rewriting the field $\hat{\varphi}_*$ (also $\hat{\varphi}$) in terms of the creation and annihilation operators, i. e.,
\begin{equation}
\hat{\tilde{\varphi}}(t,\mathbf{x})= \sum_{\mathbf{k_*}}[\hat{a}_{\mathbf{k_*}}u_{\mathbf{k_*}}(t,\mathbf{x})+ \hat{a}_{\mathbf{k_*}}^{\dagger}u_{\mathbf{k_*}}^*(t,\mathbf{x})],
\label{eq:modeexpansion}
\end{equation}
where, 
\begin{equation}
u_{\mathbf{k_*}}(t,\mathbf{x}) = \frac{1}{\sqrt{2\pi}}\frac{1}{\sqrt{2\omega_*}} e^{i(\mathbf{k_*}.\mathbf{x}-\omega_* t)}
\end{equation}
and $\mathbf{k_*}$, $\omega_*$, $\mathbf{x}$ and $t$ are dimensionless variables.

The dimensionless part of the Hamiltonian, $\hat{H}_{\rm eff}$ as given by \eqref{HTotal} and can then be expressed in the frequency space as 
\begin{equation}
\int \bigg[\frac{1}{2}\big(\frac{\partial \tilde{\varphi}}{\partial t}\big)^2+  \big(\nabla\tilde{\varphi}\big)^2 +\frac{1}{2}m_*^2\tilde{\varphi}^2\bigg]dA = \sum_{\mathbf{k_*}}(\hat{a}_{\mathbf{k_*}}^{\dagger}\hat{a}_{\mathbf{k_*}}+\frac{1}{2})\omega_* ,
\end{equation}
which leads to 
\begin{equation}
H_{\rm eff}(\varphi)= T \sum_{\mathbf{k_*}}(\hat{a}_{\mathbf{k_*}}^{\dagger}\hat{a}_{\mathbf{k_*}}+\frac{1}{2})\omega_* , 
\end{equation}
The dispersion relation is
\begin{equation}
\omega_*^2 = \mathbf{k_*}^2+ {1}/{C^2} . 
\label{Dispersion}
\end{equation}
Of course, an equivalent relation concerning $\mathbf{k}$ and $\omega$ can be written from \eqref{Dispersion}, 
\begin{equation}
T^2\omega_*^2 = T^2\mathbf{k_*}^2+ {T^2}/{C^2} . 
\end{equation}
Substituting Eq.~\eqref{deltaphiOp} in Eq. \eqref{ZetaF} leads to~\cite{1966-Kubo-RPP}: 
\begin{equation}
\zeta[\omega_{\rm IR}] = \operatorname{Im}\left[ - \frac{i}{4C^2} 
\frac{\omega_{\rm IR}}{E_\beta(\omega_{\rm IR})}\int_{-\infty}^{\infty}\langle
\hat{\varphi}(0)\hat{\varphi}(t) + \hat{\varphi}(t)\hat{\varphi}(0)
\rangle e^{-i\omega_{\rm IR} t} dt \right], 
\label{Zeta2a}
\end{equation}
where 
$E_\beta(\omega_{\rm IR})$ is the average energy of excitation in the mode with frequency $\omega_{\rm IR}$ at temperature $T$ and is given by:
\begin{equation}
\label{eq:EnergyOmega}
E_\beta(\omega_{\rm IR}) = \frac{\hbar \omega_{\rm IR}}{2} 
\coth{\left(\frac{\beta \hbar \omega_{\rm IR}}{2} \right)} \, .
\end{equation}

Now we demand that the horizon exists in the future as 
none of the physical processes can make the horizon disappear! This 
necessitates the imposition of the future boundary condition or Teleological boundary condition~\cite{1986-Price.Thorne-PRD,1986-Thorne.etal-Membrane,2016-Bhattacharya.Shankaranarayanan-PRD}. Thus, the teleological boundary condition can be viewed as a condition for the stability of the black hole event horizon. Using the teleological boundary condition, the mode expansion in \eqref{eq:modeexpansion} for $\hat{\varphi}(t)$ can be written as 
\begin{equation}
\hat{\varphi}(t)= \sum_{\omega_*'}\hat{\varphi}_{\omega_*'}(0) e^{i\omega_*' t}\theta(-t) \, . 
\label{PhiExpansion}
\end{equation}
where $\theta(-t)$ is the theta function and enforces the anti-casual, teleological nature of the horizon. (See \cite{2016-Bhattacharya.Shankaranarayanan-PRD} for a more detailed discussion.) Substituting the above expansion \eqref{PhiExpansion} in \eqref{Zeta2a}, we get, 
\begin{equation}
\zeta[\omega_{\rm IR}] = \operatorname{Im}\left[
 -\frac{i}{2C^2} \frac{\omega_{\rm IR}}{E_\beta(\omega_{\rm IR})}\sum_{\omega_*'}\frac{1}{2\pi}\langle\hat{\varphi}^2_{\omega_*'}(0)\rangle \int_{-\infty}^{\infty} dt \, \theta(-t)\exp{\left[i(\omega_*'-\omega_{\rm IR}/T)t\right]}. \right], 
 \label{Zeta3-01}
\end{equation}
Replacing the sum over $\omega_*'$ by an integral and performing the integral over $t$, we get,
\begin{equation}
\zeta[\omega_{\rm IR}] = 
\operatorname{Im}\left[ 
- \frac{i}{8 \, \pi C^2} \frac{\omega_{\rm IR}}{E_\beta(\omega_{\rm IR})}\int_{-\infty}^{\infty}\langle\hat{\varphi}^2_{\omega_*'}(0)\rangle\frac{d\omega_*'}{[-i(\omega_*'-\omega_{\rm IR}/T)]} \right], 
\label{Zeta4}
\end{equation}
where we have extended the range of the integral over negative values of $\omega'_*$ also.
Note that the above integral has a pole at $\omega_*' = \omega_{\rm IR}$, which is the well-known pole at the hydrodynamical limit. 
To evaluate the integral over $\omega_*'$, we impose the following physical condition:
\[
 \lim_{\vert\omega_*'\vert\rightarrow \infty}\langle\hat{\varphi}^2_{\omega_*'}(0)\rangle=0 \, .
\] 
This allows us to evaluate the integral over $\omega_*'$ by taking a semicircular contour on the upper half of the complex plane. The contour is constructed, so one approaches the semicircle of the contour $\vert\omega_*'\vert\rightarrow\infty$. Performing this integral, $\zeta[\omega_{\rm IR}]$ is given by:
\begin{equation}
\zeta[\omega_{\rm IR}] = 
\operatorname{Im}\left[ 
 -\frac{i}{4 C^2} \frac{\omega_{\rm IR}}{E_\beta(\omega_{\rm IR})} 
 \langle\hat{\varphi}^2_{\omega_{\rm IR}}(0)\rangle \right] \, .
 \label{Zeta5-01}
\end{equation}

Note that $\zeta[\omega_{\rm IR}]$ is evaluated at the hydrodynamic limit.
 We now have to evaluate $\langle\hat{\varphi}^2_{\omega_{\rm IR}}(0)\rangle$ for a given $\omega_{\rm IR}$.  {$\langle\hat{\varphi}^2_{\omega_{\rm IR}}(0)\rangle$ is}
\begin{equation}
\langle\hat{\varphi}^2_{\omega_{\rm IR}}(0)\rangle = \Tr\left[\frac{e^{-\beta\hat{H}_{\rm eff}}}{Z}\hat{\varphi}^2_{\omega_{\rm IR}}(0) \right] \quad  \mbox{where} \quad 
Z= \Tr[\exp{-\beta\hat{H}_{\rm eff}}] \, .
\label{Phisq}
\end{equation}

In the path-integral representation, this corresponds to evaluating the Gaussian path integral with periodic boundary conditions. This leads to:
{\small
\begin{equation}
\Tr\left[\frac{\exp{[-\beta\hat{H}_{2+1}}]}{Z}
\hat{\varphi}^2_{\omega_{\rm IR}}(0)\right]
 = \frac{\int d\tilde{\varphi}(\mathbf{k_*},\omega_{\rm IR};0)\exp\left[
 {-\frac{1}{2}m_*^2\tilde{\varphi}^2(\mathbf{k_*},\omega_{\rm IR};0)}\right] \tilde{\varphi}^2(\mathbf{k_*},\omega_{\rm IR};0)}{\int d\tilde{\varphi}(\mathbf{k_*},\omega_{\rm IR};0)\exp\left[
 {-\frac{1}{2}m_*^2\tilde{\varphi}^2(\mathbf{k_*},\omega_{\rm IR};0)}\right]} \, . 
\end{equation}
}
\begin{adjustwidth}{-\extralength}{0cm}

\reftitle{References}


\end{adjustwidth}
\end{document}